\documentclass[article]{aa}                           
\usepackage[varg]{txfonts}
\usepackage{graphicx}
\usepackage{array}
\usepackage{lscape}                                
\usepackage{txfonts}
\usepackage{caption}
\usepackage{tablefootnote}

\usepackage{natbib}
\usepackage{longtable}
\usepackage{color}
\usepackage{supertabular}
\usepackage[draft]{hyperref}
\usepackage[flushleft]{threeparttable}
\usepackage{multirow}
\usepackage{lscape}
\usepackage{comment}
\usepackage{booktabs}
\usepackage{verbatim}

\newcommand{\kms} {km\,s$^{-1}$}

\newcommand{\vsini} {$v$\,sin\,$i$}

\newcommand{\vmacro} {$v_{\rm mac}$}

\newcommand{\Teff} {$T_{\rm eff}$}
\newcommand{\grav} {log\,{\em $g$}}

\newcommand{\gravc} {log\,{\em $g_{\rm c}$}}
\newcommand{\micro} {$\xi_{\rm t}$}
\newcommand{\helio} {Y$_{\rm He}$}
\newcommand{\fastwind} {{\sc fastwind}}
\newcommand{\ioni}[2]{{#1\,\sc{#2}}}

\newcommand{\msol}{$M_{\odot}$}
\newcommand{\Lspdef} {$\mathcal{L}$}
\newcommand{\Lsp} {log\,($\mathcal{L}$/$\mathcal{L_{\odot}}$)}

\begin{document}
%
\title{The IACOB project}
\subtitle{XVI. Surface helium abundances in Galactic O-type stars: \\ indications for identifying binary interaction products
} 

\author{S.~Sim\'on-D\'iaz\inst{1,2}, G. Holgado\inst{1,2}, C. Mart\'inez-Sebasti\'an\inst{1,2}, M. Carretero-Castrillo\inst{3,4}, H. Jin\inst{5},  M.~A.~Urbaneja\inst{6}, \\ R. Gamen\inst{7},  J. Puls\inst{8}, A. de Burgos\inst{4}, M. Garcia\inst{9}, A.~Herrero\inst{2,1}, Z. Keszthelyi\inst{10}, N. Langer\inst{5,11}, F. Najarro\inst{9}, \\ J. M. Paredes \inst{3},  M. Rib\'o \inst{3}
}

\institute{
            Instituto de Astrof\'isica de Canarias, E-38200 La Laguna, Tenerife, Spain.
             \and
             Departamento de Astrof\'isica, Universidad de La Laguna, E-38205 La Laguna, Tenerife, Spain.
             \and
             Departament de Física Quàntica i Astrofísica, Institut de Ciències del Cosmos (ICCUB), Universitat de Barcelona (IEEC-UB),  c. Martí i Franquès, 1, 08028 Barcelona, Spain
            \and 
            European Southern Observatory, Alonso de C\'ordova 3107, Vitacura, Santiago, Chile
            \and
            Argelander Institut für Astronomie, Auf dem Hügel 71, DE-53121 Bonn, Germany
            \and
            Universität Innsbruck, Institut für Astro- und Teilchenphysik, Technikerstr. 25/8, A-6020 Innsbruck, Austria
             \and
            Instituto de Astrof\'isica de La Plata, Facultad de Ciencias Astron\'omicas y Geof\'isicas, CONICET--UNLP, Paseo del Bosque s/n, La Plata, Argentina.
            \and
             LMU Munich, Universitätssternwarte, Scheinerstrasse 1, 81679 München, Germany
             \and
            Centro de Astrobiología, CSIC-INTA, Crtra. de Torrejón a Ajalvir km 4, 28850 Torrejón de Ardoz (Madrid), Spain
            \and
            School of Mathematics, Statistics and Physics, Newcastle University, UK
            \and
             Max-Planck-Institut für Radioastronomie, Auf dem Hügel 69, DE-53121 Bonn, Germany
}
		   
\offprints{ssimon@iac.es}

\date{Date}

\titlerunning{Surface helium abundances in Galactic O-type stars}
\authorrunning{Sim\'on-D\'iaz et al.}

%
\abstract
{The presence of massive O-type stars with surfaces enriched by CNO-cycle products has been known since the early 1980s. For many years, internal rotational mixing was assumed to be the dominant mechanism responsible for this chemical contamination. However, accumulating evidence suggests that binary interaction may play an equally important, if not dominant, role.}
{We aim to carry out a large-scale investigation of surface helium (He) abundances in Galactic O-type stars, based on the results from the analysis of high-quality spectroscopic data from the IACOB project.}
{We perform a homogeneous spectroscopic analysis of 318 Galactic O-type stars with the {\sc iacob-broad} and {\sc fastwind/iacob-gbat} tools, deriving rotational velocities, atmospheric parameters, and He abundances. We also account for the influence of binarity, and parameter degeneracies 
on the abundance determinations.}
{We present homogeneously determined surface He abundances (\helio\,=\,N$_{\rm He}$/N$_{\rm H}$) for the so far largest, statistically significant sample of Galactic O-type stars. About 60\% of the stars show He abundances consistent with the 
cosmic abundance standard of \helio\,=\,0.098\,$\pm$\,0.002. For another 18\% of the stars, we obtain anomalously low He abundance estimates, reaching values down to 0.07. These unusual He abundances might be a consequence of flux contamination of the analyzed spectra by a faint companion. The remaining 22\% display clear He enrichment (\helio\,$\gtrsim$\,0.13). We provide observational evidence indicating that most of these He-enriched stars are likely the products of binary interaction. }
{Our study highlights how large spectroscopic surveys are gradually opening robust observational avenues to identify the products of massive binary interaction. It also emphasizes the need for caution when interpreting the spectroscopic properties of apparently single O-type stars. A significant fraction may in fact be the outcome of binary evolution rather than isolated stellar birth.}
%
%

%
\maketitle
%
%

\section{Introduction}\label{section1}

\citet{Herrero1992} carried out the first systematic quantitative spectroscopic analysis of a sample of Galactic O-type stars using plane-parallel, hydrogen+helium, non-LTE stellar atmosphere models. They reported that more than 60\% of the 25 analyzed stars exhibited helium (He) abundances higher than the cosmic standard. This result was unexpected given the understanding of stellar interiors and evolution at the time \citep[e.g.,][]{Maeder1990}, according to which the large radiative envelope surrounding the convective core of a massive main-sequence star was expected to prevent processed material produced in the core during hydrogen burning from reaching the stellar surface.

Most of the He enriched stars were either fast rotators or objects with low surface gravities and high luminosities, indicating that, from a single-star evolutionary perspective, they would have already evolved significantly away from the zero-age main sequence (ZAMS). This inconsistency between the abundances resulting from the spectroscopic analyses and theoretical predictions became widely known as the {\em helium discrepancy problem} 
This finding was in line with earlier works by \citet{Walborn1970, Walborn1971, Walborn1976}, \citet{Kudritzki1983}, \citet{Bohannan1986}, \citet{Voels1989}, and \citet{Schonberner1988}, who had already reported stars not only with He enriched surfaces, but also with carbon, nitrogen and oxygen (CNO) abundance patterns indicative of CNO-cycle processing.

\citet{Maeder1987} (see also \citealt{Langer1992}, and references therein) proposed that turbulent diffusion in the radiative zones of massive main-sequence stars, potentially triggered by differential rotation, could lead to a significant redistribution of elements during the core hydrogen burning phase. This mechanism offered a plausible explanation for the detection of stellar surfaces enriched in CNO-cycle products. Building upon this idea, a new generation of stellar interior models was developed to incorporate the effects of rotation on the evolution of massive stars \citep[see the review by][]{Maeder2000}. These models \citep[e.g.,][]{Heger2000, Brott2011, Ekstroem2012} aimed to reproduce the observed surface abundance patterns while also addressing several other discrepancies between earlier theoretical predictions and observations, including the so-called {\em mass discrepancy problem} \citep[][]{Groenewegen1989, Herrero1992}.

Several observational studies of small- to medium-sized samples of O- and B-type stars in different metallicity environments soon followed \citep[e.g.,][]{Herrero1999, Herrero2000, Repolust2004, Mokiem2005, Hunter2009, Przybilla2010, RiveroGonzalez2012, Bouret2012, Bouret2013, Bouret2021, Martins2015a, Martins2015b, Martins2016, Martins2017, Martins2024, Grin2017, Cazorla2017, Markova2018, Dufton2018, Dufton2020}. Contrary to the expectations, these works provided growing empirical evidence that rotational mixing is not the sole 
mechanism responsible for the appearance of CNO-cycle products at the surfaces of massive main-sequence stars. The detection of a non-negligible fraction of slowly rotating stars (as measured by their projected rotational velocities, \vsini) whose surfaces were N-enriched was one of the most challenging results  
\citep[][]{Morel2008, Hunter2008}.

Solutions proposed to reconcile theory with observations comprise: improved treatments of internal angular momentum transport and convective boundary mixing \citep{Maeder2014, Simoniello2015}; a reassessment of the 
reliability of spectroscopically derived abundances \citep{Maeder2014}; the influence of magnetic fields \citep[e.g.,][]{Keszthelyi2019, Keszthelyi2020}; and the inclusion of additional transport mechanisms such as internal gravity waves \citep[e.g.,][]{Aerts2014, Brinkman2025, Mombarg2025}. The role of binary interaction, in particular through mass-transfer episodes \citep[e.g.,][]{Vanbeveren1988, Vanbeveren1993, deMink2009, Song2018, Richards2024, Jin2026}, is also gathering increasing observational support.

Surface CNO abundances, together with \vsini, have been the main observational diagnostics used to test the efficiency of rotational mixing and to evaluate  alternative scenarios (see references above). 
Alternatively, \citet{Proffitt2016, Proffitt2024} used boron abundance estimates derived from UV spectra \citep[see also model predictions by][]{Frischknecht2010, HJin2024}. 

Although the study of He abundances by \citet{Herrero1992} played a key role in motivating subsequent developments in massive-star evolution models, studies specifically focused on this element remain relatively scarce \citep[e.g.,][]{Herrero2000, Repolust2004, Mokiem2005, Martins2015a, Markova2018, Aschenbrenner2023}. This is partly due to the 
complexity of deriving reliable He abundances \citep[particularly in B-type stars, but also in O-type; see, e.g.,][]{Villamariz2000, Villamariz2002, Najarro2006}. In addition, the high baseline abundance and slow surface enrichment timescale of He makes it  more difficult to detect surface variations compared to other elements.

In this paper, we embark on the first large-scale investigation of surface He abundances in Galactic O-type stars. Our study is based on the analysis of a high-quality spectroscopic dataset assembled within the framework of the IACOB project \cite{Simon-Diaz2011, Simon-Diaz2015, SimonDiaz2020a}. This work continues the efforts initiated in \citet{Holgado2018, Holgado2022} and \citet{MartinezSebastian2025}, and is complemented by a parallel study of surface N abundances (Martínez-Sebastián et al., subm.).

The structure of the paper is as follows. The description of the working sample and observational material is provided in Sect.~\ref{section2}, while Sect.~\ref{section3} concentrates on the methodology used to obtain estimates of the 
atmospheric parameters and He abundances. Section~\ref{section4} summarizes key results of our investigation, including a comparison with 
the literature and the general distribution of He abundances in several contexts.
A discussion of the results, our main conclusions, and future prospects are presented in Sects.~\ref{section5} and \ref{section6}.

\section{Sample description}\label{section2}

The sample considered in this work comprises 318 Galactic O-type stars with at least one high-resolution spectrum available in the IACOB spectroscopic database \citep[last described in][]{SimonDiaz2020a}, and which fulfill the following criteria: (1) they have not been identified as clear double-lined or higher-order spectroscopic systems; (2)  the available spectra have a signal-to-noise ratio ({\em S/N}) above 50; (3) they do not exhibit peculiar spectral features \citep[e.g., Oe, Ope, or Of?p, see][]{Sota2011, Sota2014, MaizApellaniz2016} that affect the diagnostic lines used to determine spectroscopic parameters; (4) they have both \ioni{He}{i} and \ioni{He}{ii} lines strong enough to be reliable as effective temperature indicators; and (5) they do not show a H$_{\beta}$ line in emission (i.e. hypergiants are excluded). A few additional stars where excluded from the sample after performing the quantitative spectroscopic analysis, as briefly described in Sect.~\ref{section3}. 

Our sample densely covers the region of the Hertzsprung–Russell diagram populated by O-stars, all of them being located within the main sequence band. Furthermore, the \vsini\ distribution of the sample is similar to that presented in \citet{Holgado2020} , with a main $\sim$75\% component of stars centered at $\sim$80\,\kms, and an extended high-velocity tail (\vsini\,$\gtrsim$\,200\,\kms) reaching up to $\sim$450~\kms.

Among 237 of 318 stars for which we have multi-epoch spectroscopy, 73 were identified as single-line spectroscopic binaries (SB1). This identification was based either on radial velocity variations detected across all available spectra in the IACOB database \citep[following][]{Holgado2018, Holgado2019, SimonDiaz2024}, or on a detailed investigation of spectroscopic binarity within the OWN survey \citep[][2026]{Barba2010, Barba2017}. The remaining stars were classified as likely single (LS), although some may still harbor undetected companions.

Information on the runaway (RW) status is available for $\approx$85\% of the stars in the sample. This was extracted from \citet[][]{MaizApellaniz2018} and \citet{CarreteroCastrillo2023, CarreteroCastrillo2025}, both studies based on {\em Gaia} astrometric data \citep[e.g.][]{Brown2016, Vallenari2023}. Stars for which the RW status is not available are those that did not satisfy the quality cuts established in \citet[][their Appendix A]{CarreteroCastrillo2023} to ensure reliable astrometric data.

As a final point of interest, our sample includes a significantly enhanced number of ON stars (19) compared with the 12 ON stars analyzed in the most recent comprehensive study of these type of objects by \citet{Martins2015b}.

\section{Methodology}\label{section3}

We used the best {\em S/N} spectrum per each star from the IACOB spectroscopic database\footnote{\href{https://research.iac.es/proyecto/iacob/iacobcat/}{https://research.iac.es/proyecto/iacob/iacobcat/}}. All spectra have a resolving power between $R$\,=\,25\,000 and $R$\,=\,85\,000, and typically cover the wavelength range from $\sim$3900 to 9000~\AA. The mean of the {\em S/N} distribution is 150$\pm$55, with a minimum of 50, and 80\% of the stars having {\em S/N}\,$\gtrsim$\,100.

We follow the same strategy as in  \citet{Holgado2018} to derive the line-broadening and spectroscopic parameters of the sample. Briefly, we first estimated \vsini\ and the macroturbulent broadening (\vmacro) using the {\sc iacob-broad} tool \citep{Simon-Diaz2014}, following the procedures outlined in \citet{Simon-Diaz2007, Simon-Diaz2014} and \citet{Simon-Diaz2017}. Other spectroscopic parameters, such as the effective temperature (\Teff), surface gravity (\grav), microturbulence (\micro), wind-strength $Q$ parameter \citep{Puls1996}, and surface He abundance (\helio\,=\,N$_{\rm He}$/N$_{\rm H}$), were then determined using {\sc iacob-gbat} \citep{Simon-Diaz2011, Sabin-Sanjulian2014, Holgado2018}.

\citet{Holgado2019} and \citet{Holgado2018, Holgado2020, Holgado2022} already analyzed a large fraction of the stars in our sample. However, there are two important updates with respect to the results presented there.
First, we studied 117 additional stars\footnote{A certain percentage of them not fulfilling the criteria described in Sect.~\ref{section2}, and hence excluded for the final sample under study.} for which spectra were not available at the time of those publications. Second, we reanalyzed the entire sample using an extended version of the grid of {\sc fastwind} models \citep{Santolaya-Rey1997, Puls2005, RiveroGonzalez2011} employed in \citet[][Table 2]{Holgado2020}. We optimized the new grid for this study and computed it from scratch with {\sc fastwind} v10.6.5, including: (1) a reduced step size in \helio, changed from 0.05 to 0.02, (2) a lower minimum He abundance of 0.04, and (3) an improved sampling of microturbulence below 15\,\kms,
together with two additional grid points at 25 and 30\,\kms. The full grid was computed neglecting the impact of wind-clumping.

We reanalysed the full sample with this updated grid for two main reasons:  firstly, to improve the accuracy of our He abundance estimates and, secondly, to enable a more detailed investigation of a non-negligible subsample of stars for which the original grid yielded upper limits on the He abundance of about 0.08 (further details in Sect.~\ref{lowHe}).

Table~\ref{gridFAST} summarizes the parameter space covered by the extended {\sc fastwind} grid. Table~\ref{LinesDiag} lists the complete set of optical hydrogen (\ioni{H}{i}) and helium (\ioni{He}{i}, \ioni{He}{ii}) lines synthesized in the {\sc fastwind} models. By default, in the {\sc iacob-gbat} analyses, we used the full set of indicated lines, and allowed the tool to explore a broad range of values for all free parameters, except for the velocity-law exponent, which was fixed to $\beta$\,=\,1. We also took advantage of the option to quickly recompute the best-fitting solution after excluding selected diagnostic lines or fixing specific grid parameters. This capability allowed us to assess in an objective, yet efficient, way the impact of individual lines or parameters on the derived He abundances (see Appendix~\ref{AppImpact}). 

\begin{table}[!t]
        \caption{Parameter space covered by the grid of \fastwind\ models at solar metallicity.} 
        \label{gridFAST}
        \centering
        \begin{threeparttable}
            \begin{tabular}{lcrl}
                \hline \hline
        \noalign{\smallskip}
                Parameter & Range  & Step size & Units \\
                \hline
        \noalign{\smallskip}
                \Teff\        &     22\,000 ... 55\,000  &         1\,000 & [K] \\
                \grav\        &     2.6 ... 4.4      &         0.1  & [dex] \\
                \micro\       &     1 ... 15  & 2 &  [\kms] \\
                              &     15 ... 30 & 5  &  [\kms] \\
                $Y_{\rm He}$   &     0.06 ... 0.20  & 0.02 & \\
                         & 0.20 ... 0.30 & 0.05 & \\
                log~$Q$\,$^a$     &     $-$11.9 ... $-$12.7  & 0.2 & [dex] \\
                                               &     $-$13.0 ... $-$15.0 & 0.5 & [dex] \\
                $\beta$                        &     0.8 ... 1.2  & 0.2 & \\
                \hline
        \end{tabular}
    \begin{tablenotes}
       \item \small{$^a$ $Q$\,=\,$\dot{M}$/($v_{\infty}R$)$^{1.5}$; with $\dot{M}$ in M$_{\odot}$~yr$^{-1}$, $v_{\infty}$ in \kms, $R$ in R$_{\odot}$}
     \end{tablenotes}
     \end{threeparttable}
\end{table}

\begin{table}[!t]
\caption{Diagnostic lines used in the \textsc{iacob-gbat} spectroscopic analysis of our sample of Galactic O-type stars}
\label{LinesDiag}
\centering
\begin{tabular}{cccc}
\hline \hline
\noalign{\smallskip}
\ioni{H}{} & \ioni{He}{i} & \ioni{He}{ii} & \ioni{He}{i} + \ioni{He}{ii}\\
\hline
\noalign{\smallskip}
H$\alpha$ & $\lambda$4387 & $\lambda$4200 & $\lambda$4026 \\
H$\beta$ & $\lambda$4471 & $\lambda$4541 & $\lambda$6678 + $\lambda$6683\\
H$\gamma$ & $\lambda$4713 & $\lambda$4686 & \\
H$\delta$ & $\lambda$4922 & $\lambda$5411 & \\
& $\lambda$5875 & & \\
\hline
\end{tabular}
\end{table}

As in \cite{Holgado2018}, we considered both \vsini\ and \vmacro\ were as fixed parameters in the default {\sc iacob-gbat} analyses, using the values quoted in columns 3 and 4 of Tables~\ref{tableResults_low} to \ref{tableResults_high}. We took these values directly from the outcome of the {\sc iacob-broad} analysis, with the exception of the \vmacro\ estimates for those stars with a \vsini\,$\geq$\,200\,\kms, in which case we fixed \vmacro\ to zero.

 We benefited from the ability of {\sc iacob-gbat} to provide a more complete and objective exploration of the parameter space (compared to traditional by-eye techniques). However, we adopted its results with a critical view. As with any automated method, it is essential final revision by the user. We therefore performed a visual assessment of the agreement between the best-fitting model and the observed spectrum. This step allowed us to identify cases requiring adjustments, such as refining the radial-velocity correction or modifying the adopted line-broadening parameters, as well as situations in which the derived parameters were unreliable. The latter included limitations of the {\sc fastwind} grid (e.g., the use of 1D unclumped models) or the misclassification of composite spectra as originating from single stars.

\section{Results}\label{section4}

Tables~\ref{tableResults_low} to \ref{tableResults_high} summarize the relevant information of the 318 Galactic O-type stars analyzed in this study. The stars are grouped in three tables according to their classification as He-low, He-normal, or He-rich, as defined in Sect.~\ref{distHe}. Within each group, stars are sorted first by spectral type (SpT) and then by luminosity class (LC), following the classifications reported in the Galactic O-Star Catalog \citep[GOSC\,v4.2,][]{MaizApellaniz2013, MaizApellaniz2017}. For each star, we provide the adopted values of \vsini\ and \vmacro\ (columns 3 and 4) used as input in the {\sc iacob-gbat} analysis, as well as the results of the analysis (columns 5 to 9) obtained when all free parameters were allowed to vary and the full set of diagnostic lines was considered. We also include the so-called spectroscopic luminosity, defined as \Lsp\,=\,4\,log(\Teff)\,$-$\,\grav\,$-$\,10.61 \citep{Langer2014}, as well as 
the quality flag assigned from the visual assessment of the agreement between the best-fitting model and the observed spectrum (column 10), the number of spectra available to identify whether the star is a spectroscopic binary, and indicates whether the star has been identified as a SB1 (column 12) and/or a runaway (column 13).

In this paper, we primarily focus on the He abundances resulting from the {\sc iacob-gbat} analysis. For a more detailed discussion of other aspects of the sample -- such as their physical properties (including spin rates) and the identification of potential binary interaction products -- we refer to \citet{Holgado2020, Holgado2022, Britavskiy2023, MartinezSebastian2025} and \citet{CarreteroCastrillo2025}.

\begin{figure}[!t]
\includegraphics[width=0.49\textwidth]{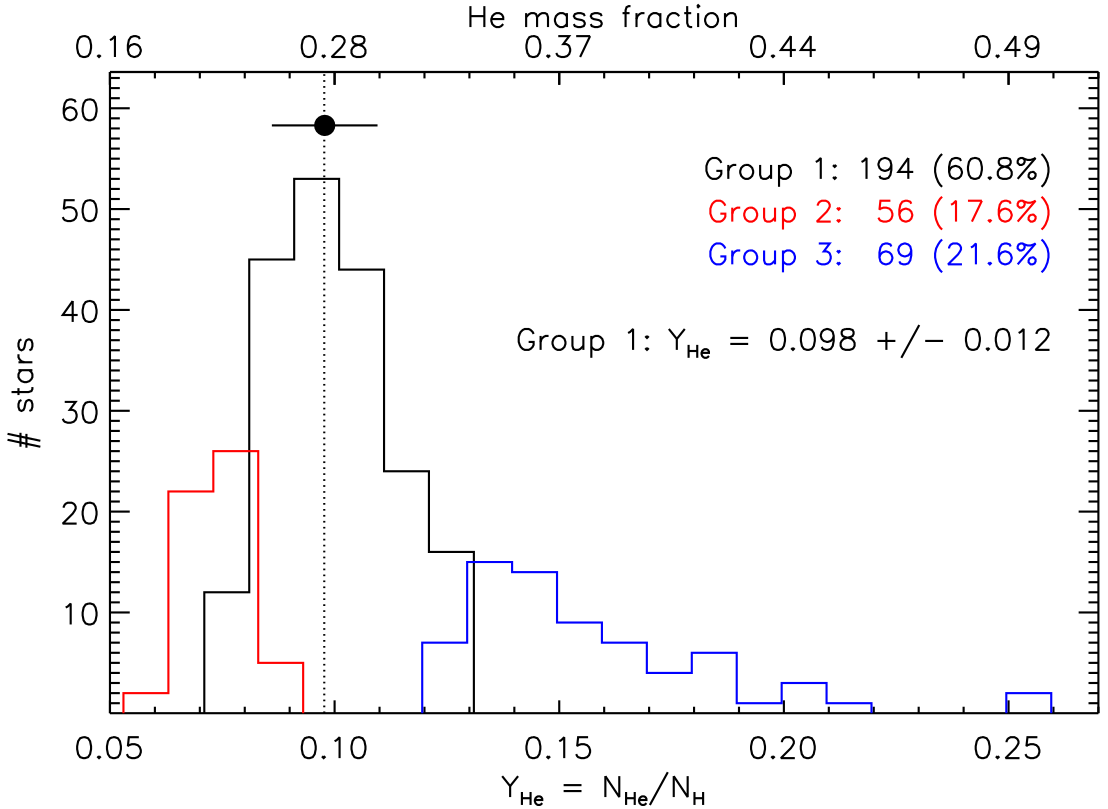}
\caption{He abundance  distributions for the three groups of stars introduced in Sect.~\ref{distHe}. Vertical dotted line indicates the present-day cosmic reference value provided by \cite{NievaPrzybilla2012}. Black dot and horizontal line indicate mean and standard deviation associated with Group\,1 stars.}
\label{Histo_YHe}
\end{figure}

\subsection{General distribution of He abundances}\label{distHe}

Figure~\ref{Histo_YHe} presents the first comprehensive distribution of surface He abundances for a statistically significant sample of Galactic O-type stars analyzed homogeneously. We divide the sample into three groups using as reference the present-day cosmic abundance of He 
\citep[\helio\,=\,0.098\,$\pm$\,0.002][]{NievaPrzybilla2012}. This reference abundance was obtained from a thorough quantitative spectroscopic analysis of a carefully selected sample of early-B type stars in the Solar Neighborhood. Group\,1 comprises 193 stars ($\sim$61\% of the sample) whose estimated abundances are compatible -- taking into account their associated uncertainties ($\Delta$\helio) -- with the indicated reference value. Group\,2 gathers the non-negligible number of 56 stars ($\sim$18\%) for which the default {\sc iacob-gbat} analysis yields unrealistically low He abundances (i.e., \helio\,+\,$\Delta$\helio\,<\,0.098, Sect.~\ref{lowHe}). Group\,3 covers the 69 stars in the high He abundance tail of the distribution ($\sim$22\%)   showing surface He enrichment compared with the reference value (i.e. \helio\,--\,$\Delta$\helio\,>\,0.098). Some overlap occurs between the values of the different Groups,
which arises from the individual uncertainties associated with stars whose He abundances fall in between $\sim$\,0.07\,--\,0.09 and $\sim$\,0.12\,--\,0.13, respectively (see Fig.~\ref{Histo_YHe}). In this regard, we note that typical (formal) uncertainties in \helio\ resulting from the default {\sc iacob-gbat} analysis for Groups\,1 to 3 are 0.025, 0.014, and 0.045 (i.e. 26, 21, and 28\%), respectively. Hereafter, we will call the stars comprising these three groups He-normal, He-low, and He-rich, respectively. Figs.~\ref{gbat_normal}, \ref{gbat_rich}, and \ref{gbat_low} in Appendix~\ref{App_examplesgbat} shows three illustrative examples of the outcome of the {\sc iacob-gbat} analysis for stars labeled as Q1 and comprising each one of these three He-abundance groups.

\subsection{Comparison with results from the literature}\label{compalit}

\begin{figure}[!t]
\includegraphics[width=0.5\textwidth]{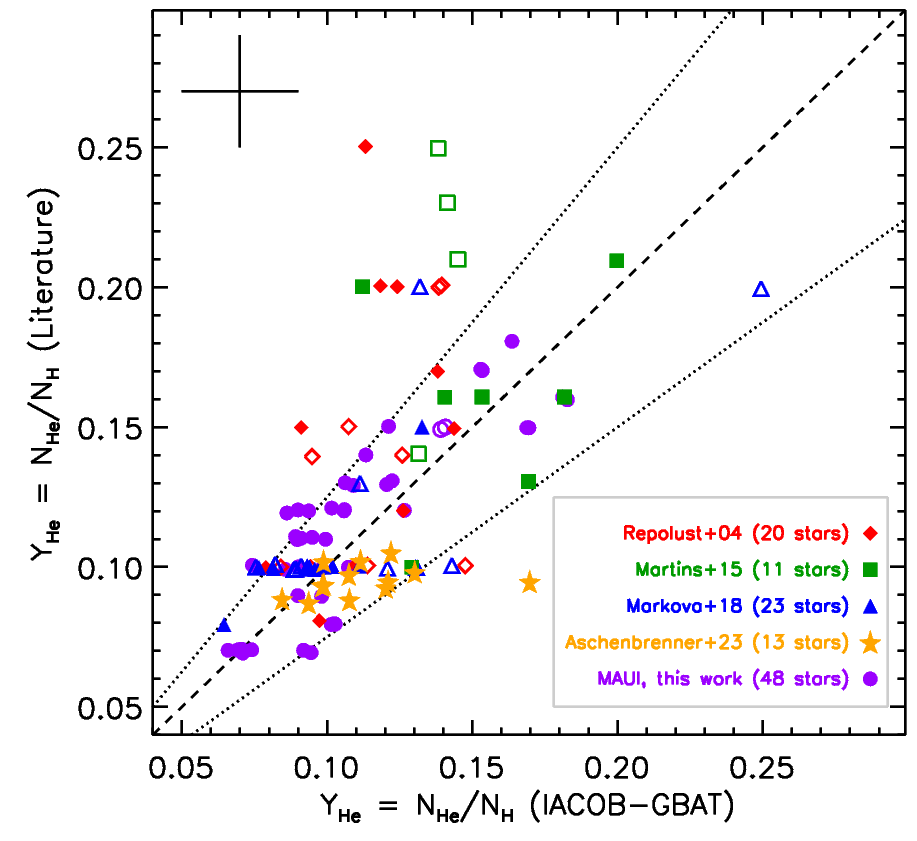}
\caption{Comparison of He abundances for a sample of 67 stars in common with the literature and 48 stars independently analyzed for this work using the code {\sc maui}. Open symbols refer to stars for which we have provided a Q2 or Q3 quality flag to the {\sc fastwind} fits (see Appendix~\ref{App_tables}). The 1-to-1 relation and 25\% tolerance region are shown as dashed and dotted lines, respectively. The cross at the top left corner indicates the typical uncertainties in the \helio\ estimates.}
\label{Fig:compalit}
\end{figure}

Figure~\ref{Fig:compalit} shows a comparison of our He abundance estimates with some others available in the literature.
We focus on four recent studies selected because they provide abundances for at least ten stars in common with our sample: namely \citet{Repolust2004,  Martins2015b, Markova2018} and \citet{Aschenbrenner2023}. The first two are based on the {\sc fastwind} atmosphere code, as in our study; the third employs {\sc cmfgen} \citep{HillierMiller1998}; and the last one relies on a hybrid analysis combining {\sc atlas9} atmosphere models \citep{Kurucz1993} with spectral synthesis calculations performed with {\sc detail} and {\sc surface} \citep[see also][]{NievaPrzybilla2012}.

Taking into account the associated uncertainties, we find a reasonably good agreement for most of the (67) stars in common with any of the abovementioned studies. Nevertheless, a small subset of (14) stars shows discrepancies beyond 25\%. Appendix~\ref{AppComp} provides
additional notes on these specific stars. 

\ \begin{figure*}[!t]
\sidecaption
\includegraphics[width=12cm]{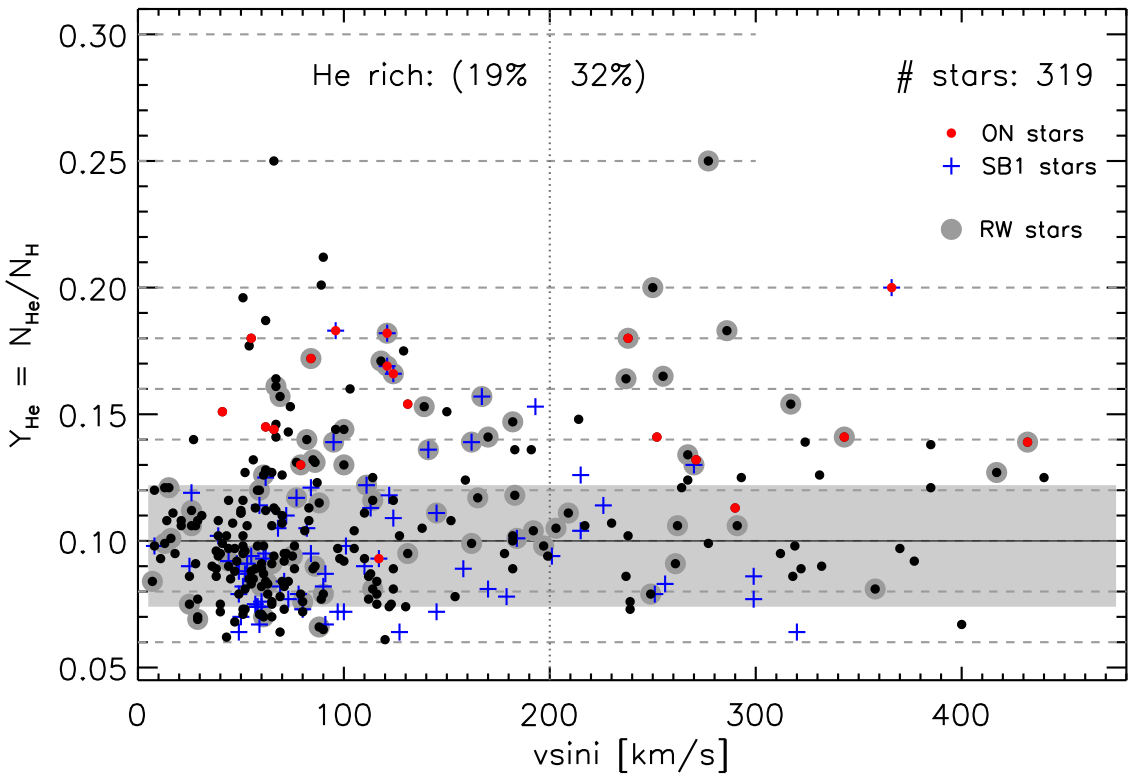}
\caption{Distribution of the 318 Galactic O-type stars in a modified Hunter diagram, using He abundance instead of nitrogen on the y-axis. The fractions of He-rich stars in the slow- and fast-rotating subsamples are indicated, assuming \vsini\,=\,200~\kms\ as the dividing threshold. ON, SB1 and RW stars are indicated with red small circles, blue crosses, and grey circles, respectively.}
\label{YHevsvsini}
\end{figure*}

We also performed a fully independent analysis of a subsample of $\sim$50 of the stars under study with the code {\sc maui} \citep{Urbaneja2026}. {\sc maui} is a modular framework that builds on a statistical emulator (in this case of {\sc fastwind} synthetic spectra) with supervised machine-learning techniques and, coupled with MCMC sampling, enables a robust and efficient spectroscopic inference for massive star parameters and surface abundances. In particular, this subsample of stars has been specifically selected to cover the full range of \Teff, \grav, \helio, and \vsini\ characterizing the complete sample analyzed with {\sc iacob-gbat}. As illustrated in Fig.~\ref{Fig:compalit}, the agreement of results between these two analysis approaches is also quite remarkable.

\subsection{The Hunter and spectroscopic HR diagrams}\label{Hunter-sHRD}

Figure~\ref{YHevsvsini} shows the distribution of stars in a modified version of the so-called Hunter diagram \citep[c.f.][]{Hunter2008}, where the He abundance is used instead of nitrogen. Stars classified as ON or identified as SB1 and/or RWs are highlighted separately. For reference, we indicate in grey the main range of He abundances (mean value $\pm$2~$\sigma$) covered by stars in Group\,1, and with grey dashed lines the step values in \helio\ defining the grid of {\sc fastwind} models incorporated into {\sc iacob-gbat}. 
We also report the percentage of He-rich stars among the samples with a \vsini\ below and above 200~\kms, respectively. This 
threshold has been adopted in several previous studies to separate the main low-\vsini\ component from the high-velocity tail in the distribution of projected rotational velocities commonly found in O stars \citep[e.g.,][]{deMink2013, Ramirez-Agudelo2013, Holgado2022, Sana2022, Britavskiy2023, CarreteroCastrillo2025}. As proposed by \citet{deMink2013}, stars with \vsini\ exceeding this threshold are most likely the products of binary interaction -- following mass-transfer or merger events -- rather than massive stars formed in isolation with such rapid rotation.

Figure~\ref{sHRD} presents the location of the stars in the three He abundance groups defined in Sect.~\ref{distHe} within a spectroscopic Hertzsprung–Russell diagram \citep[sHRD,][]{Langer2014}. As in Fig.~\ref{sHRD}, ON, SB1 and RW stars are differentiated from the rest of the sample. For reference, we also show the position of the ZAMS and the evolutionary tracks computed with the GENEC code for an initial spin rate $v_{\rm ini}$/$v_{\rm crit}$\,=\,0.4 \citep{Ekstroem2012}. Sections of the tracks where the surface He abundance is larger than 1.3 times the initial abundance are highlighted with dashed green lines. In this case, the predicted increase of the He abundance at the stellar surface is driven by the internal transport processes implemented in the models, together with the progressive removal of the outer stellar layers by winds for stars above $\sim$30\,\msol. Although we used these single star evolution tracks as reference to show a extreme case of surface enrichment due to rotational mixing (see further notes in Sect.~\ref{Hesingle}), we
note that the initial spin rate considered in these models is definitely too high when accounting from the results presented in \citet{Holgado2022}. These authors proposed that the peak of the spin distribution at birth for O-type stars is most likely located at $v_{\rm ini}$/$v_{\rm crit}$\,=\,0.10\,--\,0.15. 

\subsection{Statistical properties of the runaway and SB1 samples}\label{RWs+SB1}

\begin{table}[!t]
\caption{Summary of statistics of detected RWs and SB1 systems in the full sample and the three subsamples of He-normal/rich/low stars defined in Sect.~\ref{distHe}}
\label{Statistics_RWs_SB1}
\centering
\begin{tabular}{ccccccc}
\hline \hline
\noalign{\smallskip}
  & \multicolumn{3}{c}{Runaways} & \multicolumn{3}{c}{SB1 systems} \\
\noalign{\smallskip}
\hline
\noalign{\smallskip}
Full sample & \# & Yes & \% & \# & Yes & \% \\
\hline
\noalign{\smallskip}
All                 & 268 & 77 &  29\,\% & 237 & 73 & 31\,\% \\
\vsini\,$<$\,200    & 219 & 58 & 26\,\% & 187 & 62 & 33\,\% \\
\vsini\,$\geq$\,200 & 49 & 19 & 39\,\% & 50 &  11 & 22\,\% \\
\hline
\noalign{\smallskip}
He-normal & \# & Yes & \% & \# & Yes & \% \\
\hline
\noalign{\smallskip}
All                 & 167 & 40 & 24\,\% & 141 & 46 & 33\,\% \\
\vsini\,$<$\,200    & 138 & 31 & 22\,\% & 112 & 40 & 36\,\% \\
\vsini\,$\geq$\,200 & 29  & 9 & 31\,\% & 29 &  6 & 21\,\% \\
\hline
\noalign{\smallskip}
He-rich & \# & Yes & \% & \# & Yes & \% \\
\hline
\noalign{\smallskip}
All                 & 62 &  30 & 48\,\% & 57 &  10 & 18\,\% \\
\vsini\,$<$\,200    & 46 &  20 & 44\,\% & 41 &  8 & 20\,\% \\
\vsini\,$\geq$\,200 & 16 &  10 & 62\,\% & 16 &  2 & 12\,\% \\
\hline
\noalign{\smallskip}
He-low & \# & Yes & \% & \# & Yes & \% \\
\hline
\noalign{\smallskip}
All                 & 39 &  7 & 18\,\% & 39 & 17 & 44\,\% \\
\vsini\,$<$\,200    & 35 &  7 & 20\,\% & 34 & 14 & 41\,\% \\
\vsini\,$\geq$\,200 & 4  &  0 &  0\,\% & 5 &  3 & 60\,\% \\
\hline
\end{tabular}
\end{table}

\begin{figure}[!t]
\includegraphics[width=0.5\textwidth]{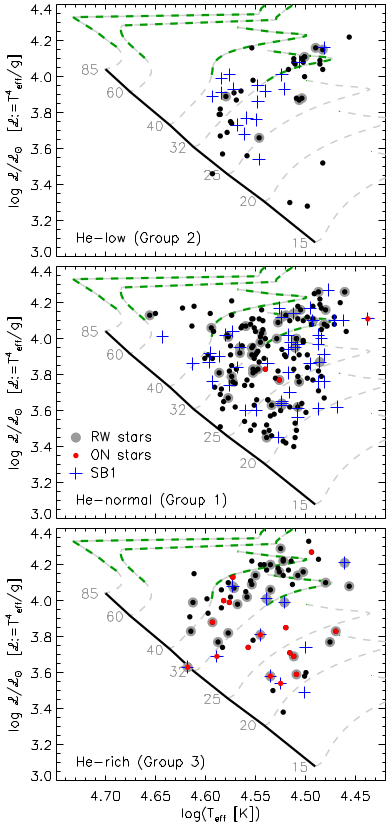}
\caption{Distribution of our sample of 318 Galactic O-type stars in a sHRD separated by the three He abundance groups described in Sect.~\ref{distHe}. Evolutionary tracks from \citet{Ekstroem2012} for an initial spin rate $v_{\rm ini}$/$v_{\rm crit}$\,=\,0.4  are depicted for reference purposes, highlighting in green the sections of the tracks where the He surface abundance reaches over 1.3 the initial abundance. Symbols are the same as in Fig.~\ref{YHevsvsini}. 
}
\label{sHRD}
\end{figure}

As stated in Sect.~\ref{section2}, we have runaway classifications for 268 stars (84.3\%) in our working sample of 318 objects. We also have sufficient multi-epoch spectroscopy\footnote{Three or more spectra.} to assess the SB1 status for 237 stars (74.5\%). In combination with their locations in the Hunter and sHR diagrams (Figs.~\ref{YHevsvsini} and \ref{sHRD}, respectively), we summarize the corresponding runaway and SB1 statistics for the full sample, as well as for subsamples defined by He abundance and \vsini, in columns 2\,--\,4 and 5\,--\,7 of Table~\ref{Statistics_RWs_SB1}, respectively.

Complementing this information, we also find that He-rich stars are much more common among runaways: 39\% of the RW sample (comprising 77 stars) are He-rich, compared to only 17\% in the non-RW sample (191 stars). Additionally, He-rich stars are less common among SB1 systems: they represent 14\% of the 73 SB1 sample, compared to 28\% among the 164 LS sample. In this regard, we remind that a certain percentage of stars identified as LS could be merger products, disrupted binaries, or undetected SB1 systems \citep[due to the still insufficient number of available epochs, or the difficulty to separate the effect of intrinsic variability from the orbital motion in a binary system where the amplitude of radial velocity variation is below 15\,--\,20~\kms,][]{SimonDiaz2024}.

\section{Discussion}\label{section5}

\subsection{The He-low sample}\label{lowHe}

One of the first aspects that drew our attention was the non-negligible fraction of stars ($\sim$18\%) for which the default {\sc iacob-gbat} analysis yielded He abundances in the range \helio\,=\,0.06\,--\,0.08 (see Fig.~\ref{Histo_YHe}). These values reach well below the lower limits of He abundances typically reported for Galactic early B-type stars \citep[e.g.,][]{Lyubimkov1975, NievaPrzybilla2012}, our Sun \citep[e.g.,][]{Serenelli2010, Moharana2024}, blue compact very metal-poor dwarf galaxies \citep[e.g.,][]{Izotov1999}, and interstellar medium estimates based on radio recombination line observations \citep[e.g.,][]{Tsivilev2023}. Such determinations have often been regarded as representative of the primordial He abundance \citep{Pagel2000}.

Given this context, it is natural to suspect that our low He abundance determinations are not physically meaningful, but rather result from limitations in the analysis. Potential contributors include incorrect estimates of microturbulence or wind properties, modeling issues in certain regions of parameter space (e.g., $T_{\rm eff}$ or $\log g$), and contamination from faint, unresolved companions whose continuum contribution can dilute the diagnostic He lines.
To investigate these possibilities, we compared stars in Groups\,2 (He-low) and 1 (He-normal) in terms of their location in the sHRD (Fig.~\ref{sHRD}) and their microturbulent velocities, as derived from the default {\sc iacob-gbat} analysis. 
We also examined whether SB1 systems are overrepresented among He-low stars (Table~\ref{Statistics_RWs_SB1}).

These comparisons reveal no significant differences in \micro\ between the two groups. In particular, the fraction of He-low stars actually decreases with increasing microturbulence, which is the opposite of what would be expected if their abundances were being systematically underestimated due to overestimated \micro. 
Likewise, although Group\,2 stars appear to cluster in a specific region of the sHRD (top panel in Fig.~\ref{sHRD}), that region also contains stars with normal He abundances, making systematic modeling limitations an unlikely explanation. Moreover, the relative percentages of He-low stars are similar among targets with assigned quality flags Q1, Q2, and Q3 (Table~\ref{tableResults_low}), again suggesting that modeling issues are not the primary cause of the anomalously low He abundances.

Altogether, these results leave as the most plausible explanation that many of the He-low stars are systems with undetected companions, for which the {\sc iacob-gbat} analysis yields spuriously low abundances. Interestingly the fraction of SB1 systems is larger in the He-low group (44\%) than in the He-normal one (33\%). Also, formal tests based on synthetic spectra computed with {\sc fastwind} -- in which the diagnostic lines were diluted by only 10\% -- show that a similar {\sc iacob-gbat} analysis to that performed here can easily underestimate the He abundance by 0.01\,--\,0.02 (Mart\'inez-Sebasti\'an et al., subm.). These tests therefore support the hypothesis that undetected companions (possibly combined with minor modeling effects) are responsible for the low He abundances found in Group\,2 stars. 

\subsection{The He-rich sample in the context of single star evolution}\label{Hesingle}

As illustrated in Fig.~\ref{Histo_YHe}, the distribution of surface He abundances is characterized by: 
(1) a dominant component comprising $\sim$80\% of the sample (Groups\,1 and 2), centered at \helio\,$\approx$\,0.095 and displaying a dispersion broadly consistent with the He abundance uncertainties of our analysis, and (2) an extended tail of He enriched stars (Group\,3), accounting for $\sim$20\% of the sample, with abundances covering the range \helio\,$\sim$\,0.12\,--\,0.25. 

Rotationally induced mixing has been considered for more than three decades
as the most likely explanation for the occurrence of these He-rich stars, with the efficiency of 
this process predicted to increase with both initial mass and rotation rate.
\citep[e.g.,][]{Maeder2000, Heger2000, Meynet2000, Brott2011, Ekstroem2012}. 
However, we provide below two independent pieces of evidence indicating that this internal mixing mechanism alone cannot explain the observed distribution of stars under study in the He\,--\,\vsini\ and sHR diagrams presented in Figs.~\ref{YHevsvsini} and \ref{sHRD}, respectively.

As in previous studies focusing on nitrogen \citep[e.g.][]{Hunter2008, RiveroGonzalez2012, Bouret2013, Grin2017}, the He\,--\,\vsini\ diagram does not reveal a clear correlation between the two quantities. Furthermore, there is a non-negligible number of He-rich stars with relatively low \vsini. In addition, while the percentage of He-rich stars within the tail of fast-rotators is   larger than in the main low-\vsini\ component (32\% vs. 19\%, see Fig.~\ref{YHevsvsini}), there is still a dominance of He-normal stars among the stars with \vsini\,>\,200~\kms.

This long-standing issue has been extensively discussed in the literature \citep{Brott2011, Maeder2014, Martins2017}, where several effects have been proposed to partially account for the observed scatter when examining individual stars. For example, age may explain the presence of some fast-rotating stars among the He-normal group, as they may simply be too young to display detectable abundance changes. This is evident from the evolutionary tracks shown in Fig.~\ref{sHRD} where, even for stars born spinning at 40\% of their critical velocity, only at the very end of the main sequence the He produced in the core is reaching the stellar surface to a detectable level. 
Similarly, the group of He-rich stars with low-\vsini\ could in principle be rapid rotators observed pole-on; however, this is unlikely given the large number of stars with these characteristics. 

Focusing only on the He\,--\,\vsini\ diagram presented in Fig.~\ref{YHevsvsini}, another plausible scenario might be that these He-rich, low \vsini\ objects provide observational evidence for efficient surface-braking mechanisms operating during the main-sequence phase, reducing the surface rotation rate while internal mixing continues to transport nuclear-processed material to the surface \citep[see, e.g.,][]{Ekstroem2012}. However,
this proposal would be in tension with the observational findings by \citet{Holgado2022}, \citet{deBurgos2024}, and \citet{Nathaniel2025} about the non detection of a clear surface braking of massive stars along their main sequence evolution.

A more critical challenge to the rotational-mixing scenario arises from the distribution of He-rich stars in the sHRD. A substantial fraction of Group\,3 stars ($\sim$47\%, i.e., those located below \Lsp\,$\sim$\,4.0) occupy regions of the diagram where the rotating GENEC tracks of \citet{Ekstroem2012} do not predict such levels of enrichment. This is   illustrated in the bottom panel of Fig.~\ref{sHRD}, where the segments of the evolutionary tracks having He abundances $>$1.3 times higher than the initial value (comparable with Group\,3 stars) are marked with thick dashed green lines. Even these models -- among the most efficient in producing surface enrichment \citep[see][]{Keszthelyi2022} -- fail to reproduce the observed population. This tension is also further strengthened by the fact that most O-type stars are likely born with initial equatorial velocities  below 0.2\,v$_{\rm crit}$ \citep{Holgado2022}, making very rapid initial rotation an unlikely explanation for the He-rich stars.

These results provide strong evidence that rotational mixing alone cannot be the dominant driver of surface He enrichment in Galactic O-type stars. Instead, our findings reinforce the growing consensus that additional mechanisms -- most notably binary interaction, as it will be shown in next sections -- might play a central role in shaping the observed He-abundance distribution.

\subsection{Some insights from binary evolution}\label{binmodels}

A large percentage of massive O stars are commonly found to be part of binary or higher-order systems \cite[e.g.][]{Kobulnicky2007, Kobulnicky2014, Chini2012, Sana2013, Moe2017, Sana2025, Mahy2009, Mahy2013, Barba2017, Offner2023, Barba2026}. \cite{Sana2012} indicated that more than 70\% of all massive stars will exchange mass with a companion at some point of their lives, leading to a binary merger in one-third of the cases. These relatively recent findings have revived a long-standing concern in stellar astrophysics -- namely, the importance of accounting for binary evolution when interpreting the observed properties of massive-star populations \citep[see, e.g., reviews by][and references therein]{Vanbeveren1988, Vanbeveren1993, Vanbeveren2004, Vanbeveren2017, Marchant2024, Marchant2026}. 

Mass transfer and merger events can modify the spin rates and surface chemical composition of the stars involved in the interaction \cite[e.g.,][]{deMink2013, Farmer2023, Menon2024, Jin2026}. Moreover, binary-interaction products will often be observed as apparently single stars \citep{deMink2014}. This will certainly be the case for merger remnants as well as runaway stars resulting from disrupted binaries following the supernova explosion of the initially more massive companion. But also for binary systems in which the post–mass-transfer donor becomes difficult to be detected, either directly in the optical spectrum or through radial velocity variations of the currently more massive and optically more luminous component, the mass gainer. Only under specific orbital configurations and mass ratios these systems will be detected as SB1.

\begin{figure}[!t]
\includegraphics[width=0.5\textwidth]{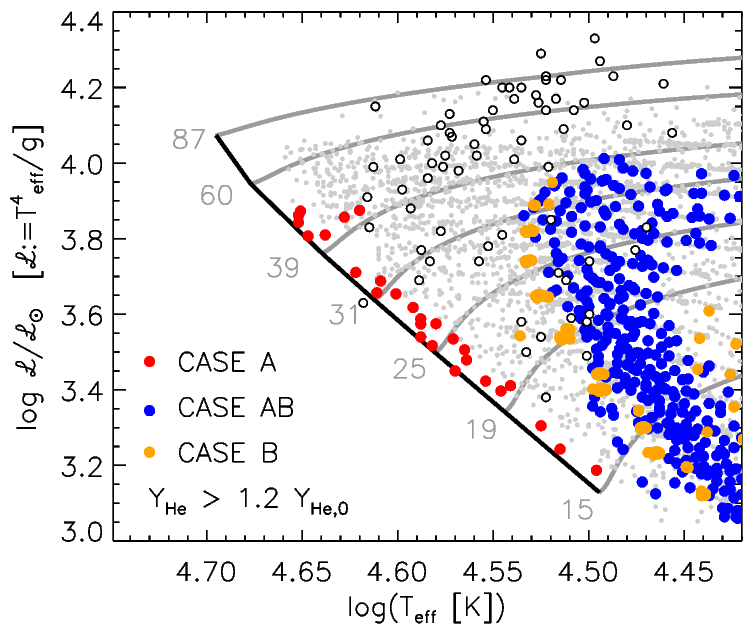}
\caption{Distribution in the sHRD of the birth location of mass gainers as predicted by the binary evolution computations by \cite{Jin2026}. The full sample of gainers is depicted with gray colors, while those with enriched He abundances are highlighted with colors separating
the mass-transfer cases A (red), B (orange) and AB (blue). Single star evolutionary tracks computed with the MESA stellar evolution code by \citet{HJin2024} and used as basis for the binary evolution computation are also depicted for reference purposes. Open circles show the location of the He-rich group of O-type stars from our study.}
\label{sHRD_bin_04vcrit}
\end{figure}

Although a detailed comparison between our observational results and theoretical predictions for binary-interaction products lies beyond the scope of this work,  we briefly comment on the expected behavior of mass gainers from binary evolution models. Figure~\ref{sHRD_bin_04vcrit} depicts the distribution in the sHRD of a sample of mass gainers as predicted by computations performed by \citet{Jin2026}. These authors have created a comprehensive grid of massive binary evolution models for solar metallicity computed with the MESA stellar evolution code \citep[first introduced in][]{Paxton2011}. They covered a range of initial primary star masses from 5 to 100 \msol. Their computations incorporate detailed stellar and binary physics, including internal differential rotation, magnetic angular momentum transport, mass-dependent overshooting, stellar wind mass-loss, mass and angular momentum transfer and tidal interaction. Specifically, the outcome of their computations presented in Fig.~\ref{sHRD_bin_04vcrit} corresponds to the moment just after mass accretion and thermal relaxation has occurred, and before further nuclear-timescale evolution has taken place. 

Within the full sample of resulting mass gainers, we highlight those with surface He abundances exceeding 1.2 times the initial value. This subsample is expected to correspond to the stars comprising Group~3, whose locations are shown again in Fig.~\ref{sHRD_bin_04vcrit} with open circles for reference. We recall that the positions of the highlighted gainers should be interpreted as their effective “birth” locations following the mass-transfer event. From these positions, all of these stars are expected to continue their subsequent evolution toward lower effective temperatures.

From inspection of Fig.~\ref{sHRD_bin_04vcrit}, and bearing in mind that He-contaminated mass gainers are expected to remain He-rich throughout their subsequent evolution \citep{Jin2026}, we can draw a general conclusion. Binary interaction through mass-transfer events can account for the presence of He-rich O-type stars in regions of the sHRD where rotating single-star evolutionary models fail to reproduce the observations (bottom panel of Fig.~\ref{sHRD}). In particular, the computations performed by \cite{Jin2026} indicate that strongly He-enriched case A mass gainers
(red filled circles) can reach the single-star ZAMS line before continuing their post mass-transfer evolution. These objects correspond to secondary stars from the highest mass binaries in the model grid, which only undergo fast case A mass transfer \citep[see section 3.1.2 in][]{Jin2026}. 
In addition, He-rich gainers resulting from case B (orange) and AB (blue) mass-transfer can also populate the region of the sHRD under study. There is a gap between the ZAMS and their positions in the diagram because the secondaries are already significantly evolved by the time the primaries deplete core hydrogen and case AB or case B mass transfer occurs. However, if mass accretion was larger than assumed in the models computed by \cite{Jin2026}, rejuvenation of this second set of gainers might be stronger and hence their birth location would be shifted towards the ZAMS line.

A second conclusion that can be extracted from inspection of Fig.~\ref{sHRD_bin_04vcrit} is that binary mass transfer does not operate efficiently at spectroscopic luminosities above \Lsp\,$\sim$\,4.0. This is because also the binary components, in particular the more massive potential donor star, partly self-strip their envelope by winds, such that much less mass is transferred if Roche Lobe overflow (RLOF) occurs. At the same time, this wind stripping keeps the binary components more compact, such that RLOF is often avoided all together \citep[see also][]{Pauli2022}. Therefore, the binary models do produce many He-enriched primary and secondary stars, in a region similar to that where He enrichment is indicated for the single star models in Fig.~\ref{sHRD}, but they do not appear in Fig.~\ref{sHRD_bin_04vcrit} as it shows only He-rich post mass transfer mass gainer models. Furthermore, additional evolutionary pathways not represented in Fig.~\ref{sHRD_bin_04vcrit} may also contribute. For instance, in so-called reverse Algol systems \citep{Sen2023}, the donor star can remain the more luminous component of the binary, exhibit surface He enrichment, and populate the upper part of the sHRD. In addition, mass accretion efficiency is one of the most uncertain parameters in binary evolution, and a more efficient mass accretion than adopted in the models of \citet{Jin2026} can help populate the upper sHRD with He-enriched mass gainers. Complementarily, although not explored here, merger events are also a viable explanation for some of the identified He-rich stars \citep{Menon2024}. 

Also remarkable the fraction of stars with He-enriched surfaces shows a clear dependence on (spectroscopic) luminosity. Below \Lsp\,$\sim$\,4.0 -- where neither rotational mixing nor wind stripping is expected to operate efficiently to enrich the surface with He -- the percentage of He-rich stars amounts to $\sim$16\%. This percentage increases to $\sim$31\% at higher luminosities, where self-stripping by winds and rotational mixing becomes more effective in both single and binary stars.

\subsection{The \helio\,--\,\vsini\ diagram}\label{section54}

The binary-evolution channel also introduces key elements for interpreting the distribution of O-type stars in the \helio\,--\,\vsini\ diagram (Fig.~\ref{YHevsvsini}), particularly in light of the shortcomings of the single star scenario (Sect.~\ref{Hesingle}). 

In systems undergoing case A mass transfer, the observed surface rotation of the He-rich gainer is not expected to exceed $\sim$200\--\,250~\kms, because tidal forces efficiently counteract the spin-up induced by mass accretion \citep[e.g.][and references therein]{deMink2013, Langer2020}. This could explain the non-negligible fraction of He-rich stars with \vsini\,<\,200~\kms\ (see Fig.~\ref{YHevsvsini} and Table~\ref{Statistics_RWs_SB1}). The same tidal effects may also explain for the relative scarcity of He-rich stars with \vsini\,$\lesssim$\,50~\kms). Additional channels -- such as reverse-Algol systems, luminous wind-stripped single and binary stars, and stellar mergers -- may also contribute to this population.

In case~B (or AB) mass-transfer events, tidal forces are no longer sufficiently strong to prevent substantial spin-up of the gainer. Such interactions can therefore produce stars that simultaneously exhibit rapid rotation and -- as shown in Sect.~\ref{binmodels} -- enhanced surface He abundances. Rotational mixing may also contribute to this population to some extent.

Finally, \citet{Jin2026} predicts that not all mass-transfer events lead to He-rich gainers. In addition, some pre-interaction binaries with low mass ratios may contribute to the He-normal population -- in many cases, but not always, detected as SB1 systems -- alongside stars evolving effectively as single.

Taken together, these effects can account for the coexistence of He-rich and He-normal stars across a wide range of projected rotational velocities and binary classifications, as observed in our sample.

\subsection{Further insights from RWs and SB1 systems}\label{section55}

Runaway stars provide valuable clues to identify past binary interactions. The peculiar velocities are generally attributed to either the binary supernova scenario \citep[BSS,][]{Blaauw1961} or to the dynamical ejection scenario \citep[DES,][]{Poveda1967}. Among these channels, the BSS is particularly relevant for understanding the surface chemical properties of massive stars. Prior to the supernova (SN), mass transfer in the binary can spin up the future RW (the gainer) and modify its chemical composition, potentially leading to He enrichment \citep{Packet1981,vandenHeuvel1985, Blaauw1993}. After the SN, the former binary will most likely result unbound, although not necessarily \citep[e.g.,][]{Renzo2019,CarreteroCastrillo2025}. These signatures -- fast rotation, altered chemical abundances, and absence of detectable companions -- are therefore expected in RW stars produced via binary interaction. 

In Sect.~\ref{RWs+SB1}, we showed that runaways are significantly more frequent among He-rich stars (48\%) than among He-normal ones (24\%), with the fraction increasing to 62\% when considering only fast rotators. This result provides strong additional support for the binary-interaction scenario as the primary explanation for the presence of He-enriched surfaces among
O-type stars, and extends the conclusions reached by \citet{Britavskiy2023} and \citet{CarreteroCastrillo2025} regarding the impact of binary interaction in these type of stars.

Conversely, the fraction of detected SB1 systems is   lower in the He-rich sample (18\%) compared to the He-normal population (33\%). Several effects can account for this reduced SB1 incidence. As discussed above, a large fraction of He-rich stars are identified as runaways, suggesting that they are mass gainers in systems that underwent mass transfer -- leading to He enrichment -- followed by disruption after the supernova explosion of the donor star (see Sect.~\ref{binmodels}). Consistent with this picture, only four out of the 30 He-rich runaway stars in our sample are detected as SB1 systems (Fig.~\ref{sHRD}).

In addition, a small fraction of He-rich stars may be merger products, in which any dynamical signature of binarity has been erased. Finally, some objects may correspond to post–mass-transfer systems that remain bound, but in which the initially more massive star has evolved into a low-mass stripped star or a compact object. In such cases, the donor becomes difficult to detect in the optical spectrum, while the mass gainer -- the currently more massive and optically more luminous component -- exhibits a relatively small orbital velocity amplitude, making SB1 detection particularly challenging. In this regard, the relatively low incidence of He-enrichment in the SB1s is consistent with the idea that binaries with significant radial velocity variations are mostly pre-interaction systems \citep{deMink2014}.

\subsection{The ON star sample}\label{ON+Herich}

From basic stellar structure physics and single-star evolution, if the observed surface abundance pattern of H-burning CNO-cycle products in (main-sequence) O-type stars were solely the result of internal mixing, He-rich stars should also display a remarkable enhancement of nitrogen at their surfaces \citep[see, e.g., Fig.~1 in][]{MartinezSebastian2025}. 
In this work, the ON qualifier used in spectral classification \citep{Walborn1970, Walborn1971, Walborn1976, Sota2011} can serve as a reliable proxy of such a strong N enrichment. These class of O-type stars were first identified by \citet{Walborn1970} as having the N lines in their optical spectra too strong for their spectral types, an anomaly which was postulated to be caused by abundance effects. This hypothesis was later confirmed by specific quantitative spectroscopic analyses \citep[e.g.,][]{Schonberner1988, Martins2015b}.

As noted in Sect.~\ref{section2}, our sample includes 19 ON stars, all of them highlighted in Figs.~\ref{YHevsvsini} and \ref{sHRD}. While ON stars are not   separated from the rest of the population in terms of \vsini\ (Fig.~\ref{YHevsvsini}), they show a strong tendency to be found among stars with surfaces enriched in He. This confirms earlier findings by \citet{Martins2015b} based on a smaller dataset.

However, the majority of He-rich stars ($\sim$80\%) are not identified as ON. Once more, a piece of observational evidence that seems to indicate that rotationally-induced mixing might not be the dominant source of contamination of the stellar surfaces in a non-negligible fraction of O-type stars. We refer the reader to \citet[][subm.]{MartinezSebastian2025} for a more detailed investigation of this result incorporating information about N abundances in a subsample of the stars considered for this work. 

\section{Conclusions and future prospects}\label{section6}

In this work we present strong observational evidence supporting the hypothesis that a large fraction of Galactic O-type stars -- classified as apparently single or SB1 systems and presenting surfaces enriched in helium -- are products of binary interaction. 

Our conclusions are grounded in the first comprehensive and homogeneous quantitative spectroscopic analysis of He abundances in a statistically significant sample of Galactic O-type stars
More than half of the stars identified as He-rich occupy regions of the sHRD that are   incompatible with the predictions of rotating single-star evolutionary models \citep[even under the most efficient internal mixing assumptions;][]{Ekstroem2012}. By contrast, these locations can be explained if these objects are mass gainers that have experienced a previous mass-transfer episode (or, in some cases, are the products of stellar mergers).

Binary-evolution models by \citet{Jin2026} predict that mass gainers resulting from case~A, B, or AB mass-transfer events can reproduce the properties of the He-rich population found along the main-sequence band spanned by typical single O-type stars (\(\sim\)15–60\,M$_\odot$; see Fig.~\ref{sHRD_bin_04vcrit}). Although not explored in detail here, some He-rich stars may also be reverse-Algol systems and merger products.  

Additional observational clues reinforce this binary-interaction interpretation. The fraction of runaways is roughly a factor of two higher among He-rich stars compared to He-normal ones, while the fraction of detected SB1 systems is somewhat lower. Both trends qualitatively agree with expectations from binary evolution, where mass gainers may become runaways following a supernova explosion, or may lose their binary signature through mergers, unbound binaries, or through the presence of optically faint stripped companions \citep{Blaauw1961, deMink2014}.

Surface He abundances hence emerge as an efficient diagnostic to identify potential binary-interaction products, providing an accessible starting point for targeted follow-up observations. More broadly, the combination of \Teff, \grav, \vsini, and He abundance measurements in statistically meaningful samples of O-type stars already offers valuable constraints for models of binary evolution and population synthesis. These constraints will become even more powerful when complemented with information on other chemical species (C, N, O), binary and runaway status, orbital parameters of SB1 systems \citep[e.g., periods and radial velocity amplitudes,][]{Barba2026}, and accurate stellar masses and luminosities \citep{Holgado2025}. The combination of all this observational information will also help to constraint to what extent the He-rich stars located above \Lsp\,$\sim$\,4.0 are also the result of binary interaction, they are produced by a combination of rotationally-induced mixing and wind-stripping, or there is combination of various of these effects.

From 318 Galactic O-type stars gathered by the IACOB project, we identified approximately 70 with clear He enrichment, corresponding to $\sim$22\% of the sample. Among them, 11 are SB1 systems whose orbital properties warrant dedicated multiwavelenght follow-up. In this sense, our study represents an intermediate step between earlier analyses of Galactic O-type stars -- typically limited to a few dozen objects -- and the order-of-magnitude increase in sample size that will soon be enabled by upcoming large-scale spectroscopic surveys such as WEAVE \citep{Jin2024} and 4MOST \citep{deJong2019}.

Finally, while this paper has focused primarily on helium, \citet[][]{MartinezSebastian2025, MartinezSebastian2026} presents a
complementary work investigating N abundances for a subset of these stars (those with \vsini\,$\lesssim$\,150~\kms). Together, these studies pave the way for a new generation of observational constraints on massive-star evolution in both single and binary channels.

\section{Data availability}
Tables xx and xx are only available in electronic form at the CDS via anonymous ftp to cdsarc.u-strasbg.fr (130.79.128.5) or via http://cdsweb.u-strasbg.fr/cgi-bin/qcat?J/A+A/.

\begin{acknowledgements}

S.S-D., G.H., C.M-S. and A.H. acknowledge support from the State Research Agency (AEI) of the Spanish Ministry of Science and Innovation (MICIN) and the European Regional Development Fund, FEDER under grants PID2021-122397NB-C21 and PID2024-159329NB-C21. 

This project received the support from the “La Caixa” Foundation (ID 100010434) under the fellowship code LCF/BQ/PI23/11970035. 

MC-C, JMP, and MR acknowledge financial support from the State
Agency for Research of the Spanish Ministry of Science and Innovation under
grants PID2022-136828NB-C41/AEI/10.13039/501100011033/ERDF/EU,
and PID2022-138172NB-C43/AEI/10.13039/501100011033/ERDF/EU, and
through the Unit of Excellence María de Maeztu 2025-2029
award to the Institute of Cosmos Sciences (CEX2024-001451-M, MICIU/AEI/10.13039/501100011033).

The project leading to this application has received funding from European Commission (EC) under Project OCEANS - Overcoming challenges in the evolution and nature of massive stars, HORIZON-MSCA-2023-SE-01, No G.A 101183150
Funded by the European Union. 

This work has made use of data from the European Space Agency (ESA) mission {\it {\em Gaia}} (\url{https://www.cosmos.esa.int/Gaia}), processed by the {\it {\em Gaia}} Data Processing and Analysis Consortium (DPAC, \url{https://www.cosmos.esa.int/web/Gaia/dpac/consortium}). Funding for the DPAC has been provided by national institutions, in particular the institutions participating in the {\it {\em Gaia}} Multilateral Agreement.

Based on observations made with the Nordic Optical Telescope, operated by NOTSA, and the Mercator Telescope, operated by the Flemish Community, both at the Observatorio del Roque de los Muchachos (La Palma, Spain) of the Instituto de Astrofísica de Canarias.  
Based on observations at the European Southern Observatory in programs 073.D-0609(A), 077.B-0348(A), 079.D-0564(A), 079.D-0564(C), 081.D-2008(A), 081.D-2008(B), 083.D-0589(A), 083.D-0589(B), 086.D-0997(A), 086.D-0997(B), 087.D-0946(A), 089.D-0975(A). 

\end{acknowledgements}

\bibliography{sample}

\appendix

\section{{\sc iacob-gbat} experiments to investigate the reliability of \helio\ estimations}\label{AppImpact}

\begin{figure*}[!t]
\includegraphics[width=\textwidth]{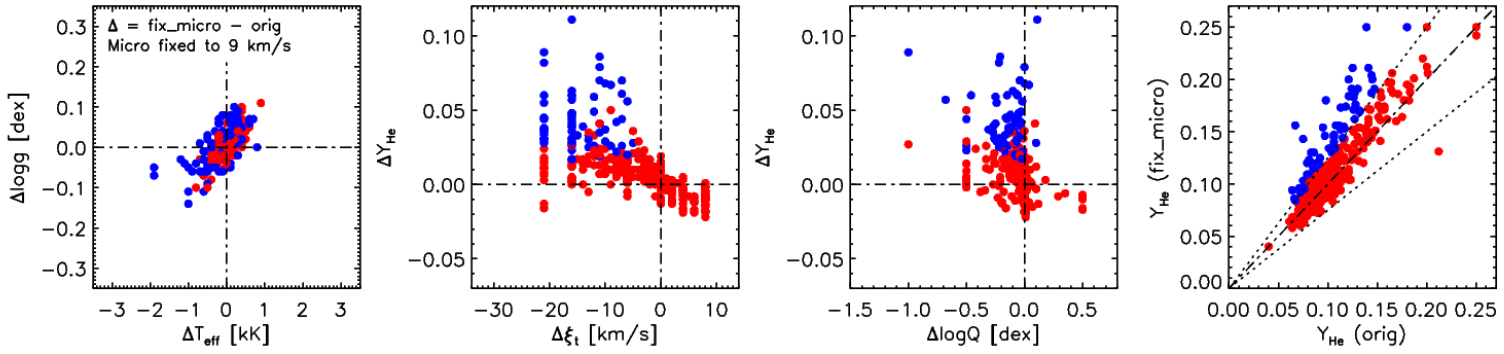}
\includegraphics[width=\textwidth]{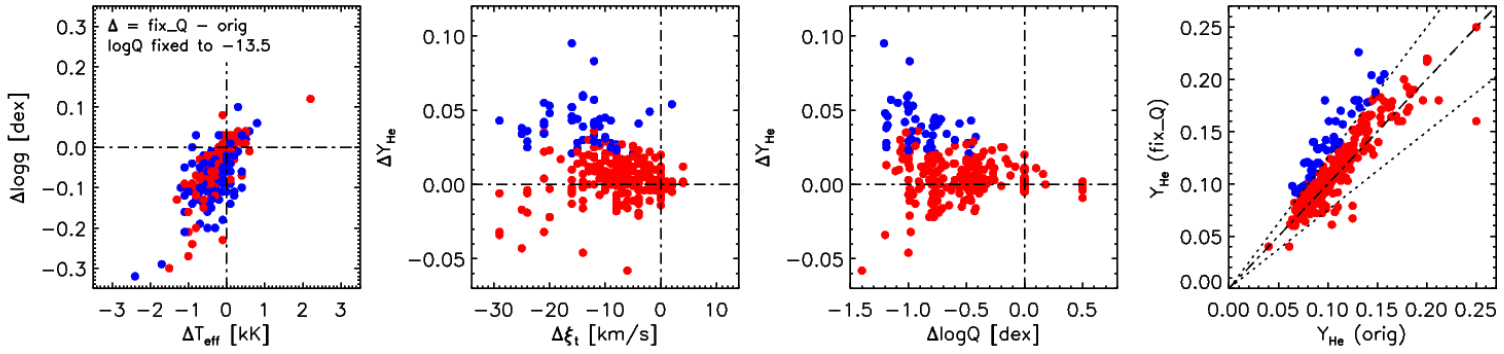}
\includegraphics[width=\textwidth]{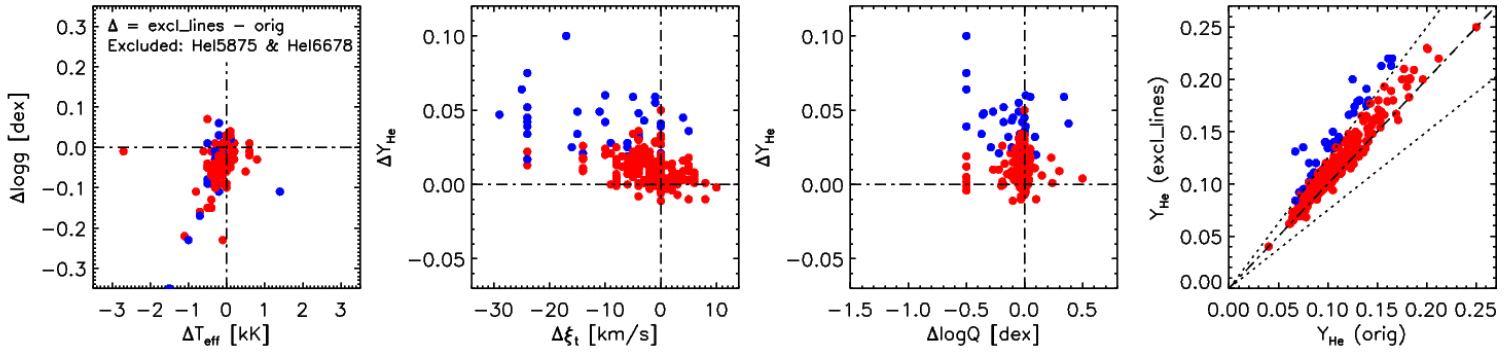}
\caption{Study of the impact of certain assumptions in the {\sc iacob-gbat} analysis}
\label{gbat_experiments}
\end{figure*}

The {\sc iacob-gbat} code provides the flexibility to perform a variety of controlled tests aimed at evaluating the impact of fixing selected fitting parameters or excluding specific diagnostic lines from the analysis. In the context of this study, we carried out three experiments of particular relevance.

The first one is motivated by the common practice in several previous spectroscopic studies of fixing the value of the microturbulent velocity (\micro) during the analysis. This approach was adopted, for example, by \citet{Martins2015a, Martins2015b}, who adopted, independently of luminosity class, a depth-variable microturbulent velocity starting from 10~\kms\ at the photosphere and reaching 10\% of the terminal velocity at the top of the atmosphere. Similarly, \citet{Repolust2004} assumed \micro = 10~\kms\ for stars later than O6 and \micro = 0~\kms\ for earlier spectral types, again irrespective of luminosity class. In the same vein, \citet{Markova2018} fixed \micro\ to 10~\kms\ for mid- and late-O stars, while adopting a value of 15~\kms\ for hotter objects.

In this first experiment, we therefore repeated the {\sc iacob-gbat} analysis using exactly the same set of diagnostic lines as in the default configuration, but fixing the microturbulence to \micro = 10~\kms. The results are summarized in the top panels of Fig.~\ref{gbat_experiments}, which illustrate the impact of this assumption on the derived values of \Teff, \grav, log$Q$, and \helio. Specifically, the three leftmost panels explore possible correlations between changes in \Teff\ and \grav, \helio\ and \micro, and \helio\ and log$Q$, respectively, while the fourth panel shows a direct comparison between the He abundances obtained in the default {\sc iacob-gbat} analysis and those resulting from this first experiment. In all panels, stars for which the difference in \helio\ exceeds 25\% are highlighted.

While the effects on the derived values of \Teff, \grav, and log$Q$ are 
generally modest, a clear and expected correlation emerges between changes in \micro\ and \helio\ (second panel from the left). In particular, we find a significant number of stars for which the He abundance increases by approximately 0.05 when adopting a fixed value of \micro\,=\,9~\kms\ in the spectroscopic analysis, with a few cases showing even larger increases of 0.07\,--\,0.10.

In the second experiment, we evaluated the impact of an incorrect determination of the wind-strength parameter, 
log$Q$. This test was motivated by the identification of several tens of stars which we have labeled with quality flag Q2 in Tables~\ref{tableResults_low} to \ref{tableResults_high} (see also Appendix~\ref{App_tables}). In these stars, despite the overall good fitting found for all diagnostic lines, we detected indications of double-peaked emission affecting the wind-sensitive H$\alpha$
and \ion{He}{ii}\,4686 lines.

To assess the potential effect of this issue, we repeated the {\sc iacob-gbat} analysis while fixing the wind-strength parameter to loq$Q$\,=\,-13.5. 
This value corresponds to a relatively weak wind and provides a representative lower limit for the objects considered in this experiment.

The results of this experiment are presented in the second row of Fig.~\ref{gbat_experiments}, using a similar set of panels as in the case of the first experiment. Again, we found a non-negligible number of cases which will result in differences in \helio\ larger than 25\%. However, this will only happen when the difference in log$Q$ is larger than 0.5 dex.

In the third and last experiment we explored what is the impact of excluding the two diagnostic lines which are more sensitive to microturbulence in this parameter domain: \ion{He}{i}\,6678 and \ion{He}{i}\,5875. These two lines are not so commonly used in other studies performing spectroscopic analyses of O-type stars, and we have found that, in the case of fast rotating stars, they seem to weight the best fitting solution (at least when using {\sc fastwind} models) towards values of \micro\ in the range 20\,--\,30~\kms\ (see Appendix.~\ref{AppComp}), independently of the parameters of the stars. As a consequence, as described in the outcome of experiment 1, this could eventually have an impact in the derived He abundances, leading to somewhat lower estimates.

The results of this third experiment, presented in the last row of Fig.~\ref{gbat_experiments}, indicates that the impact of excluding the \ion{He}{i}\,6678 and \ion{He}{i}\,5875 diagnostic lines from the {\sc iacb-gbat} analysis has a relatively small impact for about 90\% of the stars in the sample. Only 35 stars (most of them fast rotators) show differences in the derived abundances above 25\%. Interestingly, within this relatively small sample, it does not seems to be a clear correlation between modifications in the estimated \helio\ and \micro, as illustrated by the second leftmost panel in the third row of Fig.~\ref{gbat_experiments}.

Overall, the main conclusion which can be extracted from these experiments is that there might be a small percentage of stars in our sample for which we could be obtaining too low He abundances. This underestimation should be, however, very occasionally larger than $\Delta$Y$_{\rm He}$$\sim$0.05. 

\section{Stars with discrepant He abundances}\label{AppComp}

In Figure~\ref{compalit} we have presented a comparison of He abundance estimates obtained by means of the default {\sc iacob-gbat} analysis (see Sect.~\ref{section3}) and those obtained in various previous studies in the literature. This sample amounts for a total of 61 stars, including 20, 11, 23 and 13 stars in common\footnote{Note that some of the stars have been analyzed by several of the indicated authors.} with \citet{Repolust2004, Martins2015b, Markova2018} and \citet{Aschenbrenner2023}, respectively. 

As described in Sect.~\ref{compalit}, the overall agreement is quite good. However, there is a small subset of 14 stars for which we found discrepancies larger than 25\%. All these stars are quoted in Table~\ref{Notagree_Nlit}, were we also provide several information of interest, as described in the corresponding caption.

Interestingly most of the highlighted stars have been labeled with a Q2 or Q3 quality flag (App.~\ref{App_tables}). In addition, there is a quite remarkable number of them having a \vsini\ above 150~\kms. 

We are aware that there are many additional effects which can be also contributing to the identified differences in derived He abundances \citep[including, for example, the use of different spectra, some of them with poorer quality, as in the case of][or the use of different codes or diagnostic lines]{Repolust2004}. However, after inspection of Table~\ref{Notagree_Nlit}, and taking into account the results of the experiments performed in Appendix~\ref{AppImpact}, the most likely explanation is connected with discrepancies between the microturbulences assumed by previous studies and the ones determined in our analysis.

\begin{table*}[!t]
\caption{Stars in common with several studies in the literature for which we obtain discrepancies larger than 25\% in the estimated He abundances. }
\label{Notagree_Nlit}
\centering
\begin{tabular}{cccccccccccccc}
\hline \hline
\noalign{\smallskip}
 & & \multicolumn{5}{c}{This work} & &  \multicolumn{3}{c}{Literature} & & \\
\noalign{\smallskip}
\cline{3-7}\cline{9-11}
\noalign{\smallskip}
ID & SpC & \vsini & \Teff\ & \grav\ & \micro\ & \helio & &  \helio & \Teff\  & \grav\  & Qual. & SB & Ref. \\
   &     &  [\kms] &  [kK] &  [dex] &  [\kms] &  & &   &  [kK] &  [dex]  &  flag &  status &  \\
\noalign{\smallskip}
\hline
\noalign{\smallskip}
    HD\,14947  &          O4.5\,If  & 114 & 37.8 & 3.58 &   7 & 0.12 & & 0.20  & 37.5 & 3.48 &  Q1 &  LPV & R04 \\
    HD\,193682  &       O4.5\,IV(f) & 183 & 39.3 & 3.68 &   9 & 0.12 & & 0.20 & 40.0 & 3.65 &  Q1 &   LS & R04 \\ 
    HD\,210839  &   O6.5\,Iab:(n)fp & 214 & 36.3 & 3.50 &  30 & 0.15 & & 0.10 & 36.0 & 3.58 &  Q2 &   LS & R04 \\ 
    HD\,192639  &        O7.5\,Iabf &  82 & 34.3 & 3.38 &  30 & 0.14 & & 0.20 & 35.0 & 3.47 &  Q3 &    . & R04 \\ 
     HD\,24912  & O7.5\,III(n)((f)) & 230 & 35.7 & 3.51 &  25 & 0.11 & & 0.15 & 35.0 & 3.56 &  Q2 &   LS & R04 \\ 
    HD\,217086  &     O7\,Vnn((f))z & 377 & 37.5 & 3.69 &  17 & 0.09 & & 0.15 & 36.0 & 3.72 &  Q1 &   LS & R04 \\ 
     HD\,13268$^*$  &       ON8.5\,IIIn & 290 & 34.7 & 3.61 &  25 & 0.11 & & 0.25 & 33.0 & 3.48 & Q1 &   LS & R04 \\ 
     HD\,18409  &          O9.7\,Ib & 131 & 30.2 & 3.11 &  21 & 0.09 & & 0.14 & 30.0 & 3.04 &  Q2 &   LS & R04 \\ 
    HD\,191423$^*$  &     ON9\,II-IIInn & 432 & 32.5 & 3.46 &  25 & 0.14 & & 0.20 & 32.5 & 3.60 &  Q2 &   LS & R04 \\ 
\noalign{\smallskip}
     HD\,13268$^*$ &       ON8.5\,IIIn & 290 & 34.7 & 3.61 &  25 & 0.11 & & 0.20 & 32.0  & 3.63  & Q1 &   LS & M15b \\ 
    HD\,150574 &        ON\,9III(n) & 252 & 33.1 & 3.51 &  20 & 0.14 & & 0.23 & 31.0  & 3.49 & Q2 &    MD & M15b \\ 
    HD\,191423$^*$  &     ON9\,II-IIInn & 432 & 32.5 & 3.46 &  25 & 0.14 & & 0.20 & 31.5 & 3.72  &  Q2 &   LS & M15b \\ 
    HD\,123008 &         ON\,9.2Iab & 62  & 31.2 & 3.17 &  30 & 0.14 & & 0.21 & 30.0  & 3.10 &   Q3 &     . & M15b \\ 
\noalign{\smallskip}
     HD\,69464 &          O7\,Ib(f) & 73 & 35.8 & 3.38 &  25 & 0.14 & & 0.10 & 36.0  & 3.51 &    Q3 &    LS & M18 \\ 
    HD\,148546 &            O9\,Iab & 85 & 31.8 & 3.24 &  30 & 0.13 & & 0.20 & 31.0  & 3.22 &    Q3 &    LS & M18 \\ 
\noalign{\smallskip}
     HD\,14633 &           ON\,8.5V & 121 & 35.1 & 3.80 &   9 & 0.17 & & 0.10 & 34.0  & 3.9 &    Q1 &   SB1  & A23 \\
\hline
\noalign{\smallskip}
\hline
\end{tabular}
\tablefoot{
We indicate the SpC, SB status and \vsini\ of the stars, as well as the \Teff\ and \grav\ values obtained with the default {\sc iacob-gbat} analysis (see Sect.~\ref{section3} and those quoted in the reference studies (indicated in the last column). The table is complemented with the quality flag assigned to each of the {\sc iacob-gbat} analyses (see App.~\ref{App_tables} and the estimated value of microturbulence. In the case of the reference studies, this parameter was fixed to \micro\,=\,10 and 15~\kms\ for the case of dwarfs/giants and supergiants, respectively, by \citet[][R04]{Repolust2004} and \citet[][M18]{Markova2018}, and to \micro\,=\,10~\kms\ by \citet[][M15b]{Martins2015b}. The microturbulence derived by \citet[][A23]{Aschenbrenner2023} for HD\,14633 was 6~\kms.
}
\end{table*}

\section{Illustrative examples of the outcome of {\sc iacob-gbat} for stars labeled as Q1}\label{App_examplesgbat}

\begin{figure*}[!t]
\centering
\includegraphics[width=0.9\textwidth]{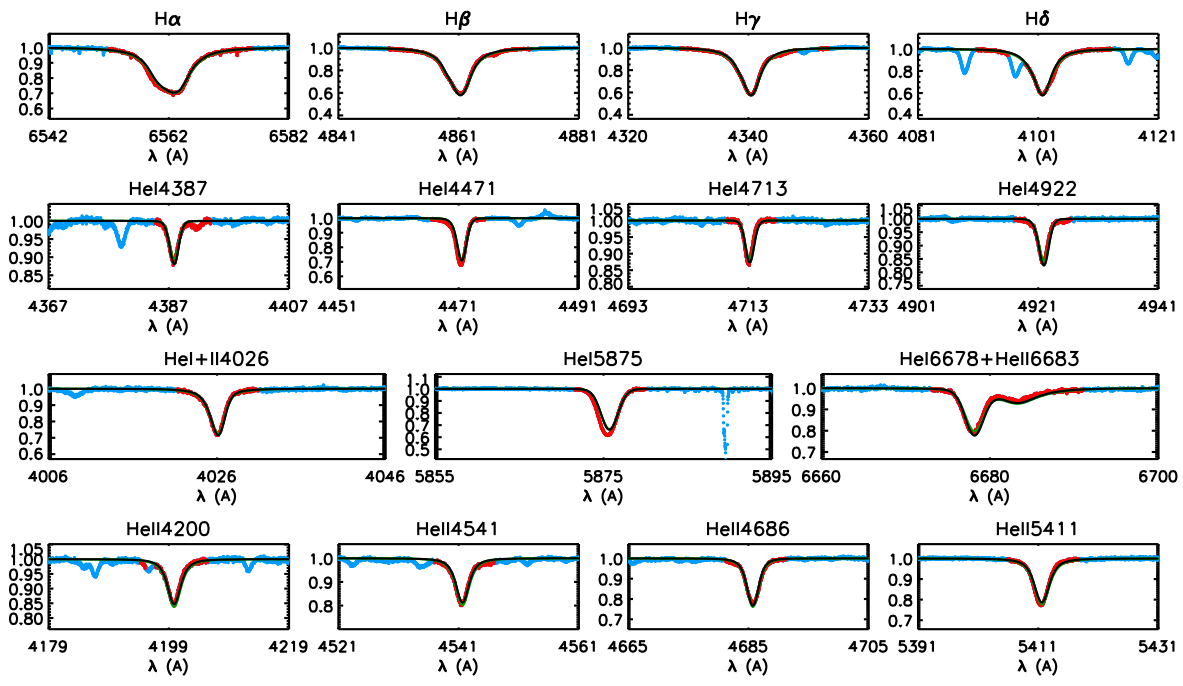}
\includegraphics[width=0.9\textwidth]{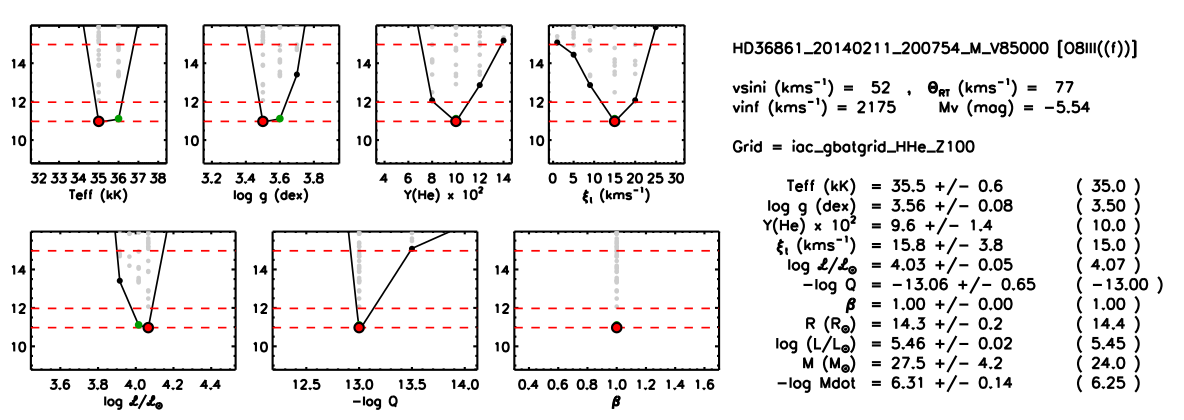}
\caption{Graphical summary of the outcome of the {\sc iacob-gbat} analysis for the O8\,III((f)) star HD\,36861, representative of the He-normal abundance group. The y-axis in the upper and lower set of panels correspond to Normalized flux and $\chi^2$, respectively.}
\label{gbat_normal}
\end{figure*}

\begin{figure*}[!t]
\centering
\includegraphics[width=0.9\textwidth]{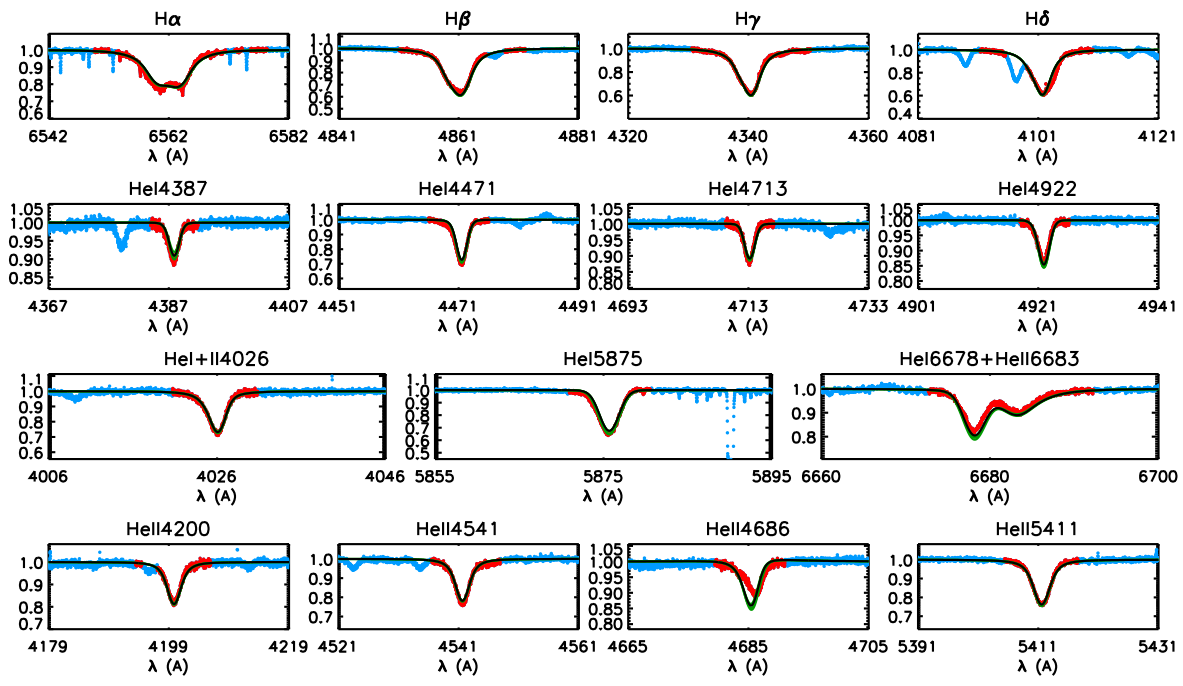}
\includegraphics[width=0.9\textwidth]{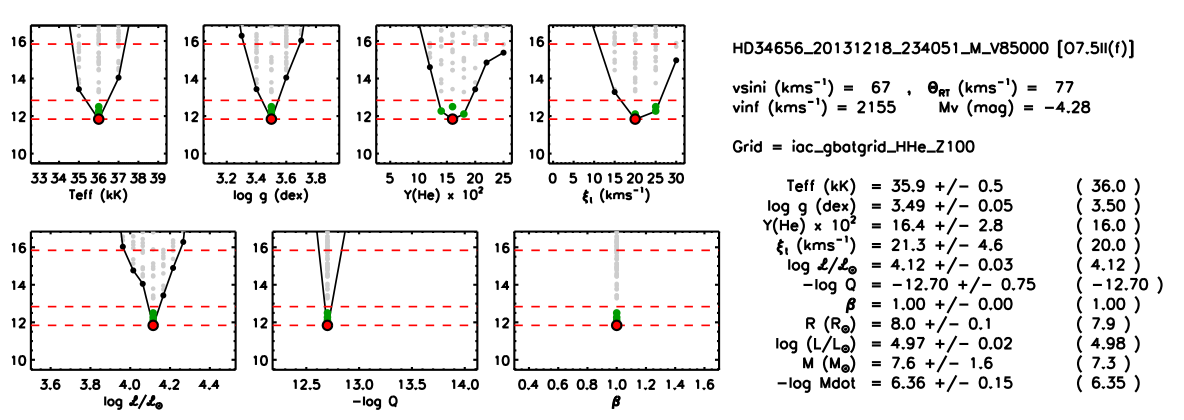}
\caption{Graphical summary of the outcome of the {\sc iacob-gbat} analysis for the O7.5\,III(f) star HD\,34656, representative of the He-rich abundance group.}
\label{gbat_rich}
\end{figure*}

\begin{figure*}[!t]
\centering
\includegraphics[width=0.9\textwidth]{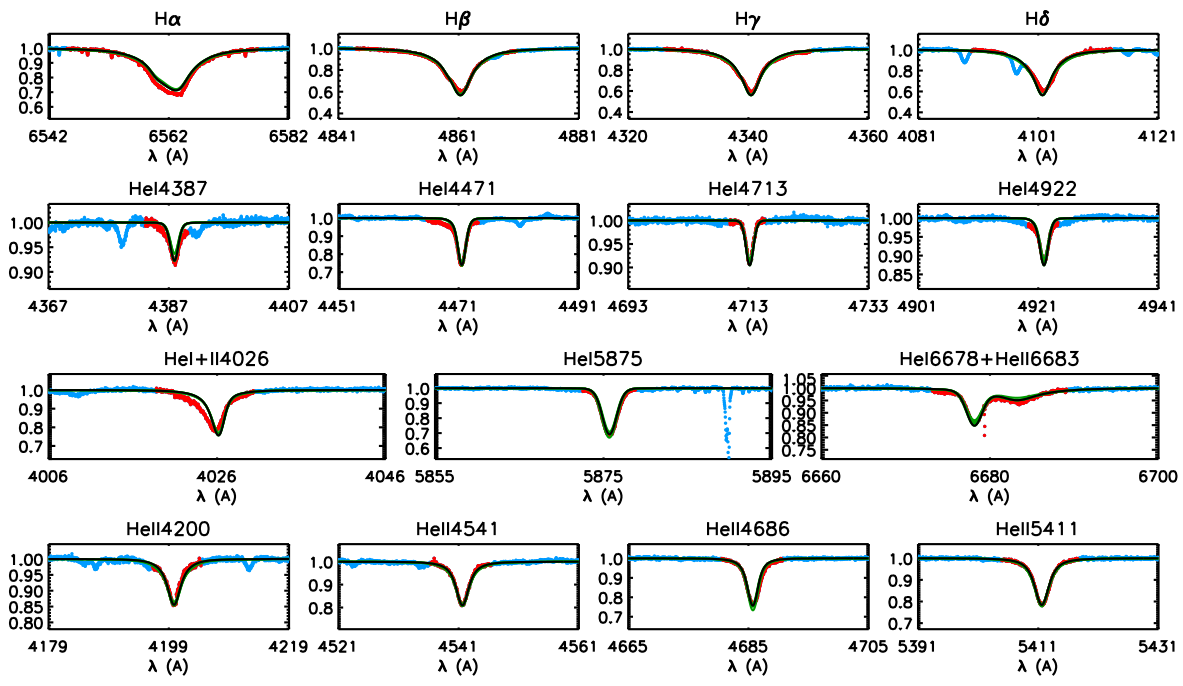}
\includegraphics[width=0.9\textwidth]{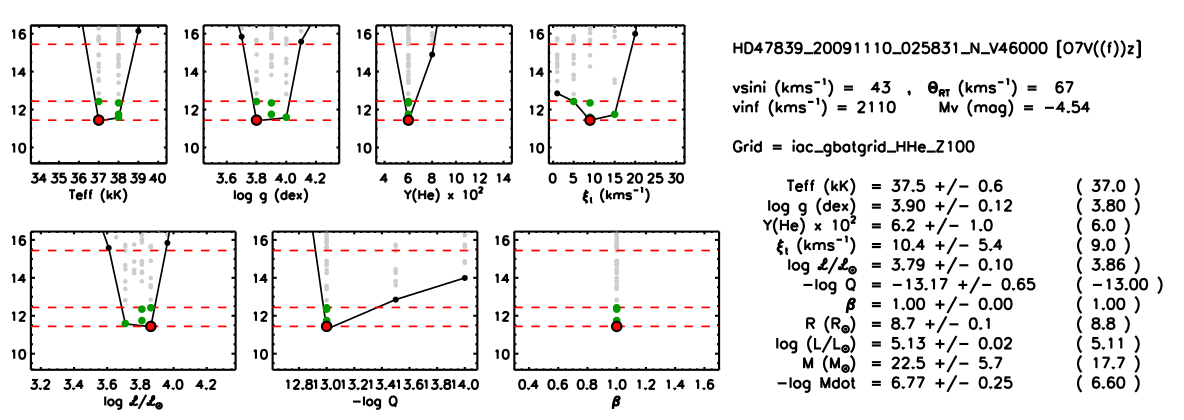}
\caption{Graphical summary of the results of the {\sc iacob-gbat} analysis for the O7\,V((f))z star HD\,47839, representative of the He-poor abundance group. Despite not being detected in the optical spectrum or through radial-velocity (RV) measurements \citep[the peak-to-peak RV dispersion measured from 109 spectra in the IACOB spectroscopic database, spanning nearly nine years, is only 3.4\,\kms, ][]{SimonDiaz2024}, HD\,47839 is known to be a long-period binary system \cite[$\sim$25\,yr, ][]{Gies1997}, including a fast-rotating B-type companion \citep[see also][]{Burssens2020}.}
\label{gbat_low}
\end{figure*}

Figures~\ref{gbat_normal}, \ref{gbat_rich}, \ref{gbat_low} depicts three illustrative examples of the high quality of the fitting to the H and He lines achieved for stars labeled as Q1. We have selected one representative star for each of the He-abundance categories defined in Sect.~\ref{distHe}.

\section{Tables}\label{App_tables}

Tables~\ref{tableResults_low}, \ref{tableResults_normal_Long}, and \ref{tableResults_high} summarize the relevant information of the 318 Galactic O-type stars analyzed in this study separated by the three groups of He abundances defined in Sect.~\ref{distHe}. See introduction of Sect.~\ref{section4} for further details.

Regarding column 10 (quality flag): Q1 refers to stars in which {\sc iacob-gbat} provides a good agreement to all considered diagnostic lines; Q2 are stars for which a good quality fit is also obtained for all diagnostic lines, but the H$\alpha$ and/or \ion{He}{ii}\,4686 lines seem to show some hits double peak emission affecting the wings of the line-profiles; lastly, the Q3 quality flag was assigned to those cases in which a simultaneous good fit to H$\alpha$ and \ion{He}{ii}\,4686 is not fulfilled.

	\onecolumn
	\pagestyle{empty}
	\fontsize{8}{8}\selectfont
	\onecolumn
	{
		\noindent\begin{longtable}{l@{\hskip 0.1in}l@{\hskip 0.1in}r@{\hskip 0.1in}r@{\hskip 0.1in}c@{\hskip 0.1in}c@{\hskip 0.1in}c@{\hskip 0.1in}c@{\hskip 0.1in}c@{\hskip 0.1in}c@{\hskip 0.1in}c@{\hskip 0.1in}c@{\hskip 0.1in}c@{\hskip 0.1in}c}
			\captionsetup{justification=raggedright, singlelinecheck=false}
			\caption{Spectroscopic parameters for the sample analyzed in this work and identified as He-low. }\\
			\multicolumn{14}{l}{\parbox{16.7cm}{Columns include: star ID; spectral classification (SpC) from the ALS catalog; projected rotational velocity ($v \sin i$) and macroturbulence (\vmacro) derived with the {\sc iacob-broad} tool; a series of parameters obtained from the {\sc iacob-gbat/fastwind} analysis including effective temperature ($T_{\rm eff}$), surface gravity corrected from centrifugal forces (\gravc) and {\bf spectroscopic luminosity \citep[\Lspdef :=\,\Teff$^4$/g, ][]{Langer2014}}; Helium abundance (\helio); microturbulence (\micro) and log\,Q. This information is complemented with a quality flag considering the fit of the {\sc fastwind} model, the number of spectra available, as well as the spectroscopic binarity (SB) and runaway (RW) status of each target.}}\\
			\label{tableResults_low}\\
			\hline \hline
			\noalign{\smallskip}
			STAR-ID & SpC & $v \sin i$ & \vmacro & \Teff & \gravc & \Lsp & \helio & \micro & logQ  & Qual. & \# sp & SB     & RW \\
			&     & [\kms]    & [\kms] &  [kK] & [dex] & [dex]  & [dex] & [\kms] & [dex] & flag  &     & status & status \\
			\hline
			\noalign{\smallskip}
			\endfirsthead
			\caption{continued.}\\
			\hline
			\hline
			\noalign{\smallskip}
			STAR-ID & SpC & $v \sin i$ & \vmacro & \Teff & \gravc & \Lsp & \helio & \micro & logQ  & Qual. & \# sp & SB     & RW \\
			&     & [\kms]    & [\kms] &  [kK] & [dex] & [dex]  & [dex] & [\kms] & [dex] & flag  &     & status & status \\
			\hline
			\noalign{\smallskip}
			\endhead
			\noalign{\smallskip}
			\noalign{\smallskip}
			\hline\hline
			\noalign{\smallskip}
			\multicolumn{14}{l}{SB status: \textbf{LS}: Likely single, \textbf{LPV}: Line profile variable, \textbf{SB1}: Single-lined spectroscopic binary.}\\
			\multicolumn{14}{l}{Classifications based on fewer than three spectra (as indicated in the \# sp. column) should be treated with caution.}
			\endfoot
BD+453216A & O5V((f))z & 69 & 57 & 36.2 ± 0.8 & 4.06 ± 0.10 & 3.56 ± 0.10 & 0.064 ± 0.010 & 1 & -13.5 & Q1.5 & 4 & LPV & . \\
CPD-592600 & O6V((f)) & 127 & 106 & 39.2 ± 1.1 & 3.87 ± 0.12 & 3.89 ± 0.13 & 0.064 ± 0.010 & 11 & -12.7 & Q1.7 & 8 & SB1 & NO \\
HD227018 & O6.5V((f))z & 61 & 78 & 38.0 ± 1.0 & 3.81 ± 0.10 & 3.89 ± 0.10 & 0.070 ± 0.013 & 14 & -13.1 & Q1.0 & 4 & LS & YES \\
HD344784 & O6.5V((f))z & 65 & 70 & 38.5 ± 0.9 & 3.92 ± 0.12 & 3.82 ± 0.13 & 0.076 ± 0.017 & 7 & -13.1 & Q1.0 & 1 & . & . \\
HD91572 & O6.5V((f))z & 60 & 57 & 38.4 ± 0.7 & 3.81 ± 0.10 & 3.91 ± 0.10 & 0.076 ± 0.017 & 1 & -13.0 & Q1.0 & 5 & SB1 & NO \\
HD242935 & O6.5V((f))z & 29 & 64 & 38.3 ± 1.1 & 4.16 ± 0.20 & 3.57 ± 0.20 & 0.070 ± 0.015 & 5 & -13.1 & Q1.0 & 1 & . & NO \\
BD+61411 & O6.5V(n)((f))z & 29 & 98 & 38.5 ± 1.2 & 3.96 ± 0.17 & 3.79 ± 0.18 & 0.077 ± 0.020 & 5 & -13.5 & Q1.0 & 1 & . & NO \\
HD93146 & O7V((f)) & 60 & 67 & 38.4 ± 0.7 & 3.73 ± 0.05 & 3.99 ± 0.05 & 0.074 ± 0.014 & 1 & -12.9 & Q1.0 & 10 & SB1 & . \\
HD47839 & O7V((f))z & 43 & 67 & 37.5 ± 0.6 & 3.97 ± 0.08 & 3.74 ± 0.08 & 0.062 ± 0.010 & 10 & -13.2 & Q1.0 & 174 & LS & . \\
HD97966 & O7V((f))z & 33 & 59 & 39.2 ± 1.1 & 4.34 ± 0.10 & 3.46 ± 0.10 & <0.040 ± 0.020 & 18 & -12.9 & Q1.0 & 1 & . & . \\
HD242926 & O7V(n)z & 89 & 52 & 38.5 ± 0.9 & 4.06 ± 0.15 & 3.67 ± 0.15 & 0.077 ± 0.015 & 1 & -13.5 & Q1.0 & 3 & LS & . \\
CPD-262711 & O7Vz & 90 & 55 & 37.9 ± 1.2 & 4.01 ± 0.21 & 3.69 ± 0.21 & 0.065 ± 0.017 & 5 & -13.5 & Q1.0 & 1 & . & NO \\
HD35619 & O7.5V((f)) & 40 & 60 & 37.6 ± 0.7 & 3.95 ± 0.09 & 3.75 ± 0.09 & 0.072 ± 0.013 & 12 & -13.1 & Q1.0 & 10 & LPV/ & . \\
HD344777 & O7.5V & 71 & 54 & 36.7 ± 0.9 & 3.69 ± 0.05 & 3.92 ± 0.05 & 0.073 ± 0.015 & 18 & -13.2 & Q1.0 & 4 & LPV & NO \\
HD53975 & O7.5V(n)z & 179 & 45 & 37.0 ± 0.6 & 3.93 ± 0.05 & 3.73 ± 0.06 & 0.078 ± 0.011 & 8 & -13.5 & Q1.2 & 6 & SB1 & NO \\
HD101413 & O8V & 80 & 91 & 36.4 ± 0.8 & 3.97 ± 0.09 & 3.68 ± 0.09 & 0.073 ± 0.015 & 1 & -13.1 & Q1.0 & 5 & SB1 & NO \\
HD66788 & O9V & 29 & 44 & 35.2 ± 0.8 & 3.93 ± 0.16 & 3.66 ± 0.17 & 0.069 ± 0.014 & 9 & -13.2 & Q1.0 & 2 & . & YES \\
HD306097 & O9V & 123 & 0 & 32.9 ± 0.5 & 3.36 ± 0.05 & 4.04 ± 0.07 & 0.075 ± 0.017 & 16 & -12.8 & Q1.7 & 1 & . & . \\
HD101070 & O9.7V & 59 & 63 & 32.8 ± 0.9 & 4.13 ± 0.17 & 3.30 ± 0.17 & 0.071 ± 0.012 & 1 & -13.5 & Q1.0 & 1 & . & . \\
HD217035 & O9.7V(n) & 116 & 50 & 31.5 ± 0.8 & 4.12 ± 0.12 & 3.28 ± 0.13 & 0.084 ± 0.013 & 1 & -13.5 & Q1.7 & 3 & LS & NO \\
HD124314 & O6IV(n)((f))+O.. & 256 & 0 & 37.3 ± 0.8 & 3.76 ± 0.06 & 3.93 ± 0.07 & 0.083 ± 0.013 & 17 & -12.6 & Q2.3 & 15 & SB1 & NO \\
HD76556 & O6IV(n)((f))p & 239 & 0 & 38.6 ± 0.9 & 3.93 ± 0.11 & 3.79 ± 0.11 & 0.073 ± 0.014 & 25 & -12.6 & Q2.3 & 4 & LS & NO \\
HD97319 & O7.5IV((f)) & 51 & 77 & 35.7 ± 0.7 & 3.61 ± 0.11 & 3.98 ± 0.11 & 0.071 ± 0.013 & 23 & -12.8 & Q1.0 & 1 & . & NO \\
HD168444 & O8IV & 50 & 73 & 35.3 ± 0.7 & 3.62 ± 0.05 & 3.95 ± 0.05 & 0.070 ± 0.013 & 15 & -13.1 & Q1.0 & 4 & SB1 & NO \\
HD52533 & O8.5IVn & 299 & 0 & 35.2 ± 0.5 & 4.03 ± 0.09 & 3.54 ± 0.09 & 0.077 ± 0.011 & 1 & -13.1 & Q1.3 & 10 & SB1 & NO \\
HD73882 & O8.5IV & 145 & 58 & 35.4 ± 0.8 & 3.82 ± 0.11 & 3.76 ± 0.11 & 0.072 ± 0.013 & 1 & -13.0 & Q1.7 & 8 & SB1 & NO \\
HD113659 & O9IV & 58 & 75 & 33.4 ± 0.7 & 3.46 ± 0.09 & 4.01 ± 0.10 & 0.074 ± 0.014 & 13 & -13.1 & Q1.0 & 5 & SB1 & NO \\
HD152314 & O9IV & 49 & 70 & 33.2 ± 0.6 & 3.54 ± 0.07 & 3.93 ± 0.07 & 0.064 ± 0.016 & 22 & -12.6 & Q3.0 & 4 & SB1 & NO \\
HD190991 & O9.2IV & 47 & 51 & 32.3 ± 1.3 & 3.58 ± 0.16 & 3.88 ± 0.16 & 0.068 ± 0.018 & 9 & -13.0 & Q1.0 & 2 & . & NO \\
HD193117 & O9.5IV(n) & 60 & 93 & 31.7 ± 0.8 & 3.28 ± 0.07 & 4.11 ± 0.08 & 0.071 ± 0.016 & 22 & -12.8 & Q1.0 & 4 & LS & NO \\
HD190427 & O9.7IV & 25 & 58 & 31.9 ± 1.1 & 3.52 ± 0.15 & 3.88 ± 0.16 & 0.075 ± 0.017 & 14 & -13.6 & Q1.0 & 2 & . & YES \\
\noalign{\smallskip}
HD152723 & O6.5III(f) & 73 & 100 & 37.7 ± 0.7 & 3.69 ± 0.06 & 4.01 ± 0.06 & 0.077 ± 0.012 & 5 & -12.7 & Q1.5 & 13 & SB1 & . \\
HD156738 & O6.5III(f) & 65 & 103 & 37.3 ± 1.0 & 3.78 ± 0.13 & 3.91 ± 0.13 & 0.070 ± 0.016 & 15 & -12.9 & Q1.5 & 2 & . & NO \\
HD93160 & O7III((f)) & 120 & 64 & 37.1 ± 0.9 & 3.81 ± 0.12 & 3.87 ± 0.13 & 0.061 ± 0.010 & 9 & -12.5 & Q2.2 & 5 & LS & . \\
HD163800 & O7.5III((f)) & 70 & 70 & 35.8 ± 0.5 & 3.42 ± 0.05 & 4.16 ± 0.05 & 0.084 ± 0.010 & 24 & -12.8 & Q1.0 & 4 & LS & NO \\
HD114737 & O8.5III & 58 & 83 & 35.3 ± 0.8 & 3.72 ± 0.05 & 3.86 ± 0.05 & 0.074 ± 0.014 & 1 & -13.0 & Q1.0 & 4 & SB1 & . \\
HD344863 & O8.5/9III & 130 & 54 & 32.5 ± 0.6 & 3.34 ± 0.04 & 4.08 ± 0.05 & 0.074 ± 0.016 & 16 & -12.9 & Q1.7 & 4 & LS & NO \\
HD215806 & O9III & 51 & 67 & 32.4 ± 0.7 & 3.35 ± 0.05 & 4.07 ± 0.05 & 0.073 ± 0.014 & 17 & -12.9 & Q1.0 & 3 & LS & NO \\
HD229234 & O9III & 97 & 53 & 32.1 ± 1.0 & 3.32 ± 0.06 & 4.08 ± 0.07 & 0.072 ± 0.018 & 11 & -13.0 & Q1.0 & 5 & SB1 & NO \\
HD171201 & O9III & 116 & 73 & 31.8 ± 0.7 & 3.32 ± 0.05 & 4.08 ± 0.06 & 0.079 ± 0.017 & 25 & -12.7 & Q1.2 & 1 & . & . \\
HD113606 & O9.5III & 40 & 81 & 30.4 ± 0.6 & 3.29 ± 0.08 & 4.04 ± 0.09 & 0.075 ± 0.017 & 11 & -13.0 & Q1.0 & 1 & . & NO \\
HD189957 & O9.7III & 88 & 54 & 32.1 ± 0.4 & 3.54 ± 0.06 & 3.87 ± 0.06 & 0.066 ± 0.010 & 13 & -13.1 & Q1.0 & 5 & LS & YES \\
HD161807 & O9.7IIInn & 400 & 0 & 30.4 ± 1.5 & 3.78 ± 0.14 & 3.52 ± 0.17 & 0.067 ± 0.025 & 30 & -13.0 & Q2.3 & 3 & LS & NO \\
HD190429B & O9.5II-III & 112 & 83 & 31.7 ± 0.6 & 3.40 ± 0.05 & 4.00 ± 0.05 & 0.077 ± 0.020 & 12 & -13.1 & Q1.2 & 3 & LS & NO \\
HD101205 & O7II:(n) & 320 & 0 & 36.2 ± 0.8 & 3.86 ± 0.07 & 3.77 ± 0.07 & 0.064 ± 0.010 & 11 & -12.5 & Q2.3 & 3 & SB1 & . \\
\noalign{\smallskip}
HD100444 & O9II & 52 & 77 & 32.5 ± 0.7 & 3.34 ± 0.07 & 4.10 ± 0.08 & 0.073 ± 0.014 & 23 & -12.7 & Q1.0 & 3 & LS & NO \\
HD328856 & O9.7II & 80 & 105 & 30.6 ± 0.7 & 3.28 ± 0.09 & 4.07 ± 0.09 & 0.072 ± 0.016 & 15 & -12.7 & Q3.0 & 1 & . & . \\
HD209975 & O9Ib & 63 & 77 & 31.9 ± 0.3 & 3.31 ± 0.05 & 4.10 ± 0.05 & 0.079 ± 0.010 & 21 & -12.8 & Q1.5 & 80 & LS & NO \\
HD155756 & O9Ibp & 62 & 99 & 31.7 ± 0.5 & 3.30 ± 0.05 & 4.11 ± 0.05 & 0.084 ± 0.010 & 25 & -12.5 & Q1.5 & 2 & . & YES \\
HD218915 & O9.2Iab & 63 & 88 & 30.9 ± 0.4 & 3.20 ± 0.05 & 4.16 ± 0.05 & 0.083 ± 0.010 & 30 & -12.6 & Q3.0 & 5 & LS & YES \\
HD37742 & O9.2IbNwk & 122 & 97 & 30.6 ± 0.6 & 3.29 ± 0.10 & 4.04 ± 0.09 & 0.074 ± 0.016 & 20 & -12.6 & Q3.2 & 177 & LPV & . \\
HD152424 & OC9.2Ia & 59 & 66 & 30.3 ± 0.8 & 3.15 ± 0.10 & 4.16 ± 0.10 & 0.067 ± 0.017 & 25 & -12.3 & Q3.0 & 5 & SB1 & NO \\
HD152003 & O9.7IabNwk & 65 & 83 & 30.6 ± 0.6 & 3.25 ± 0.09 & 4.09 ± 0.10 & 0.075 ± 0.017 & 17 & -12.5 & Q3.0 & 2 & . & NO \\
HD165319 & O9.7Ib & 80 & 101 & 30.4 ± 0.8 & 3.15 ± 0.07 & 4.15 ± 0.08 & 0.076 ± 0.019 & 20 & -12.6 & Q3.0 & 4 & LS & YES \\
HD194280 & OC9.7Iab & 116 & 67 & 28.6 ± 0.9 & 2.96 ± 0.04 & 4.22 ± 0.05 & 0.075 ± 0.020 & 19 & -12.7 & Q3.2 & 10 & LS & NO \\
HD96670 & O8.5(n)fpvar & 91 & 107 & 34.7 ± 1.1 & 3.61 ± 0.15 & 3.92 ± 0.16 & 0.067 ± 0.023 & 9 & -12.5 & Q2.0 & 8 & SB1 & . \\
     \end{longtable}
}

\onecolumn
\pagestyle{empty}
\fontsize{8}{8}\selectfont
\onecolumn
{
		\noindent\begin{longtable}{l@{\hskip 0.1in}l@{\hskip 0.1in}r@{\hskip 0.1in}r@{\hskip 0.1in}c@{\hskip 0.1in}c@{\hskip 0.1in}c@{\hskip 0.1in}c@{\hskip 0.1in}c@{\hskip 0.1in}c@{\hskip 0.1in}c@{\hskip 0.1in}c@{\hskip 0.1in}c@{\hskip 0.1in}c}
		\captionsetup{justification=raggedright, singlelinecheck=false}
		\caption{Spectroscopic parameters for the sample analyzed in this work and identified as He-normal.}\\
		\multicolumn{14}{l}{\parbox{16.7cm}{Same columns as Table~\ref{tableResults_low}}}\\
		\label{tableResults_normal_Long}\\
		\hline \hline
		\noalign{\smallskip}
			STAR-ID & SpC & $v \sin i$ & \vmacro & \Teff & \gravc & \Lsp & \helio & \micro & logQ  & Qual. & \# sp & SB     & RW \\
			&     & [\kms]    & [\kms] &  [kK] & [dex] & [dex]  & [dex] & [\kms] & [dex] & flag  &     & status & status \\
		\hline
		\noalign{\smallskip}
		\endfirsthead
		\caption{continued.}\\
		\hline
		\hline
		\noalign{\smallskip}
			STAR-ID & SpC & $v \sin i$ & \vmacro & \Teff & \gravc & \Lsp & \helio & \micro & logQ  & Qual. & \# sp & SB     & RW \\
			&     & [\kms]    & [\kms] &  [kK] & [dex] & [dex]  & [dex] & [\kms] & [dex] & flag  &     & status & status \\
		\hline
		\noalign{\smallskip}
		\endhead
		\noalign{\smallskip}
		\noalign{\smallskip}
		\hline\hline
		\noalign{\smallskip}
			\multicolumn{14}{l}{SB status: \textbf{LS}: Likely single, \textbf{LPV}: Line profile variable, \textbf{SB1}: Single-lined spectroscopic binary.}\\
			\multicolumn{14}{l}{Classifications based on fewer than three spectra (as indicated in the \# sp. column) should be treated with caution.}
		\endfoot
HD64568 & O3V((f*))z & 75 & 53 & 47.2 ± 1.8 & 3.96 ± 0.12 & 4.13 ± 0.10 & 0.094 ± 0.019 & 30 & -12.8 & Q1.5 & 6 & LS & YES \\
HD93128 & O3.5V((fc))z & 58 & 56 & 48.1 ± 2.5 & 4.04 ± 0.22 & 4.08 ± 0.20 & 0.10 ± 0.03 & 25 & -12.7 & Q1.5 & 3 & LS & NO \\
HD93129B & O3.5V((f)) & 66 & 62 & 46.7 ± 1.8 & 3.93 ± 0.11 & 4.14 ± 0.09 & 0.094 ± 0.019 & 25 & -12.7 & Q1.5 & 1 & . & NO \\
HD46223 & O4V((f)) & 51 & 112 & 42.0 ± 0.6 & 3.71 ± 0.05 & 4.17 ± 0.05 & 0.095 ± 0.015 & 25 & -12.9 & Q1.5 & 8 & LS & NO \\
HD96715 & O4V((f))z & 59 & 86 & 45.3 ± 1.8 & 3.87 ± 0.17 & 4.13 ± 0.19 & 0.12 ± 0.03 & 30 & -12.8 & Q1.5 & 2 & . & YES \\
HD164794 & O4V((f))z & 62 & 95 & 44.0 ± 1.0 & 3.88 ± 0.12 & 4.01 ± 0.13 & 0.095 ± 0.017 & 25 & -12.8 & Q1.5 & 22 & SB1 & . \\
HD15629 & O4.5V((fc)) & 76 & 64 & 41.1 ± 1.3 & 3.81 ± 0.05 & 4.07 ± 0.05 & 0.098 ± 0.023 & 9 & -12.7 & Q1.5 & 5 & LS & NO \\
HD319699 & O5V((fc)) & 69 & 76 & 41.0 ± 1.3 & 3.98 ± 0.11 & 3.86 ± 0.12 & 0.082 ± 0.020 & 5 & -12.7 & Q1.5 & 4 & SB1 & NO \\
HD93204 & O5.5V((f)) & 110 & 93 & 39.1 ± 1.0 & 3.72 ± 0.05 & 4.04 ± 0.05 & 0.093 ± 0.017 & 17 & -12.8 & Q1.7 & 3 & LS & NO \\
HD14434 & O5.5IVnn(f)p & 417 & 0 & 38.6 ± 1.3 & 3.94 ± 0.11 & 3.81 ± 0.12 & 0.13 ± 0.03 & 30 & -12.9 & Q2.3 & 6 & LS & YES \\
CPD-582611 & O6V((f))z & 39 & 73 & 39.6 ± 1.0 & 3.88 ± 0.08 & 3.91 ± 0.08 & 0.11 ± 0.03 & 15 & -13.1 & Q1.0 & 1 & . & NO \\
HD303311 & O6V((f))z & 47 & 61 & 39.9 ± 1.0 & 3.90 ± 0.11 & 3.89 ± 0.11 & 0.097 ± 0.019 & 12 & -13.1 & Q1.0 & 1 & . & NO \\
HD92206B & O6V((f))z & 238 & 0 & 37.6 ± 0.8 & 3.73 ± 0.09 & 3.96 ± 0.10 & 0.102 ± 0.024 & 17 & -13.1 & Q1.8 & 9 & LPV & NO \\
BD+62424 & O6.5V(n)((f)) & 60 & 65 & 37.8 ± 1.0 & 3.68 ± 0.09 & 4.01 ± 0.09 & 0.082 ± 0.022 & 9 & -13.0 & Q1.0 & 3 & LS & YES \\
HD199579 & O6.5V(n)((f))z & 52 & 90 & 39.5 ± 0.9 & 3.91 ± 0.11 & 3.87 ± 0.11 & 0.087 ± 0.017 & 12 & -13.1 & Q1.0 & 193 & SB1 & NO \\
HD305532 & O6.5V((f))z & 51 & 53 & 39.2 ± 1.3 & 3.88 ± 0.13 & 3.89 ± 0.13 & 0.098 ± 0.025 & 25 & -12.9 & Q1.0 & 1 & . & NO \\
ALS12370 & O6.5Vnn((f)) & 440 & 0 & 39.0 ± 2.0 & 4.03 ± 0.19 & 3.73 ± 0.20 & 0.12 ± 0.05 & 25 & -13.0 & Q1.8 & 4 & LS & NO \\
HD227465 & O7V((f)) & 61 & 79 & 37.1 ± 1.3 & 3.84 ± 0.16 & 3.83 ± 0.17 & 0.09 ± 0.03 & 15 & -13.1 & Q1.0 & 3 & LS & NO \\
ALS12619 & O7V((f))z & 21 & 43 & 38.2 ± 0.9 & 4.07 ± 0.15 & 3.65 ± 0.15 & 0.11 ± 0.03 & 8 & -14.0 & Q1.0 & 1 & . & NO \\
BD+622078 & O7V((f))z & 38 & 38 & 38.5 ± 1.0 & 4.05 ± 0.09 & 3.69 ± 0.09 & 0.10 ± 0.03 & 11 & -13.5 & Q1.0 & 3 & LS & NO \\
HD91824 & O7V((f))z & 51 & 48 & 39.2 ± 0.7 & 4.00 ± 0.09 & 3.76 ± 0.09 & 0.082 ± 0.016 & 8 & -13.0 & Q1.0 & 9 & SB1 & NO \\
HD93222 & O7V((f))z & 50 & 90 & 36.6 ± 0.7 & 3.57 ± 0.07 & 4.07 ± 0.07 & 0.111 ± 0.023 & 10 & -13.1 & Q1.0 & 3 & LS & NO \\
HD227245 & O7V(n)((f))z & 47 & 71 & 37.6 ± 0.9 & 3.79 ± 0.06 & 3.91 ± 0.06 & 0.088 ± 0.018 & 19 & -13.1 & Q1.0 & 3 & LS & NO \\
BD+60501 & O7V(n)((f))z & 182 & 87 & 37.7 ± 1.2 & 3.88 ± 0.14 & 3.82 ± 0.15 & 0.10 ± 0.03 & 9 & -13.5 & Q1.2 & 3 & LS & NO \\
HD46485 & O7V((f))n & 322 & 0 & 36.4 ± 0.8 & 3.88 ± 0.06 & 3.74 ± 0.07 & 0.089 ± 0.021 & 9 & -13.5 & Q1.3 & 83 & LS & NO \\
BD+60513 & O7Vn & 319 & 0 & 36.2 ± 1.3 & 3.85 ± 0.14 & 3.77 ± 0.16 & 0.10 ± 0.03 & 15 & -13.4 & Q1.3 & 5 & LS & NO \\
BD-104682 & O7Vn((f)) & 358 & 0 & 37.5 ± 1.6 & 3.90 ± 0.14 & 3.79 ± 0.15 & 0.081 ± 0.025 & 25 & -13.0 & Q1.8 & 2 & . & YES \\
HD217086 & O7Vnn((f))z & 377 & 0 & 37.5 ± 1.0 & 3.85 ± 0.07 & 3.85 ± 0.07 & 0.092 ± 0.023 & 17 & -13.0 & Q1.8 & 9 & LS & NO \\
HD36879 & O7V(n)((f)) & 209 & 0 & 36.8 ± 0.6 & 3.75 ± 0.04 & 3.92 ± 0.04 & 0.111 ± 0.019 & 23 & -12.8 & Q2.3 & 7 & LS & YES \\
HD46573 & O7.5V((f)) & 77 & 81 & 36.8 ± 0.7 & 3.71 ± 0.05 & 3.95 ± 0.05 & 0.117 ± 0.022 & 18 & -13.1 & Q1.0 & 5 & SB1 & YES \\
HD99546 & O7.5V((f))Nstr & 50 & 60 & 36.6 ± 0.7 & 3.69 ± 0.06 & 3.96 ± 0.06 & 0.112 ± 0.019 & 14 & -13.1 & Q1.0 & 1 & . & NO \\
HD168504 & O7.5V & 61 & 89 & 37.0 ± 1.1 & 3.82 ± 0.11 & 3.84 ± 0.12 & 0.093 ± 0.020 & 9 & -13.5 & Q1.0 & 3 & SB1 & NO \\
BD+60586 & O7.5Vz & 38 & 77 & 38.3 ± 0.7 & 4.12 ± 0.16 & 3.61 ± 0.16 & 0.089 ± 0.016 & 3 & -13.5 & Q1.0 & 5 & LPV & . \\
HD152590 & O7.5Vz & 48 & 56 & 37.7 ± 0.7 & 3.97 ± 0.09 & 3.74 ± 0.09 & 0.091 ± 0.017 & 15 & -13.0 & Q1.0 & 10 & SB1 & NO \\
HD164492 & O7.5Vz & 39 & 54 & 38.4 ± 0.8 & 4.01 ± 0.13 & 3.72 ± 0.13 & 0.094 ± 0.019 & 9 & -13.5 & Q1.0 & 15 & LS & NO \\
HD338916 & O7.5Vz & 26 & 45 & 37.8 ± 0.8 & 4.03 ± 0.11 & 3.67 ± 0.11 & 0.11 ± 0.03 & 8 & -14.0 & Q1.0 & 1 & . & YES \\
HD44811 & O7.5Vz & 26 & 43 & 37.4 ± 0.7 & 3.93 ± 0.12 & 3.76 ± 0.12 & 0.106 ± 0.018 & 9 & -13.5 & Q1.0 & 5 & LS & YES \\
BD+331025 & O7.5V(n)z & 178 & 10 & 38.5 ± 1.6 & 4.22 ± 0.28 & 3.51 ± 0.30 & 0.095 ± 0.022 & 5 & -13.5 & Q1.2 & 1 & . & NO \\
HD164536 & O7.5V(n) & 237 & 0 & 37.2 ± 0.7 & 4.05 ± 0.10 & 3.62 ± 0.10 & 0.086 ± 0.017 & 8 & -13.5 & Q1.3 & 8 & LS & . \\
HD168461 & O7.5V((f))Nstr & 192 & 30 & 35.7 ± 1.2 & 3.79 ± 0.17 & 3.83 ± 0.18 & 0.10 ± 0.03 & 15 & -12.7 & Q2.2 & 1 & . & YES \\
HD101223 & O8V & 55 & 63 & 35.4 ± 0.6 & 3.67 ± 0.07 & 3.92 ± 0.07 & 0.088 ± 0.015 & 9 & -13.1 & Q1.0 & 5 & LS & NO \\
HD145217 & O8V & 49 & 90 & 35.2 ± 1.1 & 3.71 ± 0.17 & 3.87 ± 0.18 & 0.079 ± 0.024 & 9 & -13.1 & Q1.0 & 3 & LS & . \\
HD191978 & O8V & 57 & 82 & 35.7 ± 0.8 & 3.81 ± 0.06 & 3.81 ± 0.06 & 0.089 ± 0.017 & 11 & -13.5 & Q1.0 & 4 & LS & YES \\
HD97848 & O8V & 41 & 77 & 35.3 ± 0.5 & 3.66 ± 0.08 & 3.93 ± 0.08 & 0.099 ± 0.020 & 5 & -13.5 & Q1.0 & 2 & . & NO \\
HD305438 & O8Vz & 18 & 41 & 37.2 ± 0.5 & 3.92 ± 0.06 & 3.75 ± 0.06 & 0.095 ± 0.015 & 6 & -13.5 & Q1.0 & 1 & . & NO \\
HD101191 & O8V & 138 & 101 & 35.6 ± 1.1 & 3.72 ± 0.06 & 3.89 ± 0.06 & 0.10 ± 0.04 & 15 & -13.0 & Q1.2 & 4 & LS & NO \\
ALS7833 & O8Vz & 105 & 75 & 35.5 ± 0.9 & 3.70 ± 0.07 & 3.91 ± 0.07 & 0.097 ± 0.023 & 9 & -13.5 & Q1.2 & 1 & . & NO \\
HD165246 & O8V(n) & 251 & 0 & 36.3 ± 0.8 & 4.02 ± 0.10 & 3.60 ± 0.10 & 0.079 ± 0.019 & 11 & -14.0 & Q1.3 & 18 & SB1 & NO \\
HD46056 & O8Vn & 370 & 0 & 35.6 ± 0.9 & 3.97 ± 0.08 & 3.63 ± 0.09 & 0.097 ± 0.025 & 15 & -13.5 & Q1.3 & 7 & LS & NO \\
ALS15196 & O8.5V & 66 & 41 & 36.4 ± 0.8 & 4.09 ± 0.13 & 3.54 ± 0.13 & 0.11 ± 0.03 & 1 & -13.1 & Q1.0 & 2 & . & NO \\
HD216532 & O8.5V(n) & 199 & 54 & 35.6 ± 0.8 & 4.08 ± 0.10 & 3.52 ± 0.10 & 0.094 ± 0.019 & 7 & -14.0 & Q1.2 & 25 & LS & NO \\
HD92504 & O8.5V(n) & 182 & 63 & 34.5 ± 0.7 & 3.91 ± 0.20 & 3.66 ± 0.21 & 0.102 ± 0.024 & 1 & -13.5 & Q1.2 & 4 & LS & NO \\
HD298429 & O8.5V & 105 & 80 & 33.4 ± 0.8 & 3.58 ± 0.15 & 3.94 ± 0.16 & 0.10 ± 0.03 & 18 & -13.5 & Q1.2 & 1 & . & NO \\
BD+364145 & O8.5V(n) & 203 & 0 & 35.3 ± 1.3 & 3.75 ± 0.18 & 3.83 ± 0.19 & 0.10 ± 0.04 & 11 & -13.5 & Q1.3 & 8 & LS & YES \\
HD214680 & O9V & 14 & 43 & 35.4 ± 0.6 & 3.93 ± 0.09 & 3.66 ± 0.09 & 0.108 ± 0.018 & 9 & -14.0 & Q1.0 & 39 & LS & NO \\
HD216898 & O9V & 44 & 57 & 35.9 ± 1.0 & 4.13 ± 0.18 & 3.49 ± 0.18 & 0.097 ± 0.022 & 3 & -13.5 & Q1.0 & 4 & . & NO \\
CPD-592551 & O9V & 124 & 92 & 34.5 ± 0.6 & 3.88 ± 0.08 & 3.66 ± 0.08 & 0.116 ± 0.018 & 10 & -13.5 & Q1.2 & 2 & . & NO \\
HD46660 & O9V & 124 & 122 & 32.4 ± 1.4 & 4.00 ± 0.19 & 3.45 ± 0.21 & 0.081 ± 0.023 & 1 & -13.0 & Q2.2 & 1 & . & NO \\
HD44597 & O9.2V & 15 & 34 & 34.2 ± 1.1 & 3.88 ± 0.19 & 3.63 ± 0.19 & 0.12 ± 0.04 & 5 & -14.0 & Q1.0 & 4 & LS & YES \\
HD46202 & O9.2V & 11 & 38 & 34.8 ± 0.7 & 4.08 ± 0.10 & 3.48 ± 0.10 & 0.093 ± 0.018 & 5 & -13.5 & Q1.0 & 5 & LS & . \\
HD58465A & O9.2V & 44 & 27 & 34.3 ± 1.4 & 3.94 ± 0.26 & 3.57 ± 0.27 & 0.09 ± 0.03 & 7 & -13.5 & Q1.0 & 2 & . & NO \\
HD95275 & O9.2V & 25 & 40 & 33.9 ± 0.4 & 3.72 ± 0.12 & 3.78 ± 0.12 & 0.086 ± 0.013 & 9 & -13.5 & Q1.0 & 2 & . & NO \\
BD+60499 & O9.5V & 17 & 38 & 34.0 ± 1.1 & 3.97 ± 0.18 & 3.58 ± 0.19 & 0.11 ± 0.03 & 5 & -13.5 & Q1.0 & 1 & . & NO \\
CPD-546791 & O9.5V & 31 & 39 & 34.6 ± 0.9 & 4.06 ± 0.12 & 3.52 ± 0.12 & 0.11 ± 0.03 & 9 & -13.3 & Q1.0 & 1 & . & NO \\
HD305536 & O9.5V & 60 & 35 & 34.2 ± 1.0 & 4.04 ± 0.14 & 3.46 ± 0.14 & 0.089 ± 0.020 & 5 & -13.5 & Q1.0 & 1 & . & NO \\
HD34078 & O9.5V & 13 & 32 & 34.6 ± 0.6 & 4.11 ± 0.11 & 3.45 ± 0.12 & 0.099 ± 0.022 & 5 & -14.0 & Q1.0 & 89 & LS & YES \\
HD38666 & O9.5V & 111 & 56 & 33.4 ± 0.7 & 3.83 ± 0.05 & 3.64 ± 0.05 & 0.12 ± 0.03 & 1 & -13.5 & Q1.2 & 10 & SB1 & YES \\
HD256035 & O9.5Vn & 239 & 0 & 34.4 ± 1.8 & 4.04 ± 0.26 & 3.49 ± 0.27 & 0.08 ± 0.03 & 1 & -13.5 & Q1.8 & 1 & . & NO \\
BD-134930 & O9.7V & 8 & 28 & 33.1 ± 0.9 & 4.04 ± 0.26 & 3.42 ± 0.27 & 0.098 ± 0.019 & 4 & -13.5 & Q1.0 & 4 & LPV & NO \\
HD192039 & O9.7V & 16 & 38 & 32.2 ± 0.7 & 3.78 ± 0.05 & 3.62 ± 0.06 & 0.101 ± 0.020 & 7 & -14.0 & Q1.0 & 4 & LS & YES \\
HD326329 & O9.7V & 87 & 39 & 32.3 ± 0.9 & 3.92 ± 0.05 & 3.49 ± 0.05 & 0.12 ± 0.03 & 5 & -13.0 & Q1.0 & 9 & LS & NO \\
HD54879 & O9.7V & 8 & 8 & 32.6 ± 0.8 & 4.10 ± 0.11 & 3.34 ± 0.10 & 0.12 ± 0.03 & 3 & -13.0 & Q1.0 & 8 & LS & NO \\
HD165132 & O9.7V & 127 & 20 & 32.4 ± 1.2 & 3.90 ± 0.17 & 3.53 ± 0.18 & 0.10 ± 0.03 & 9 & -14.0 & Q1.2 & 1 & . & NO \\
HD239729 & O9.7V & 100 & 27 & 32.5 ± 0.7 & 4.04 ± 0.13 & 3.40 ± 0.14 & 0.092 ± 0.016 & 7 & -13.5 & Q1.2 & 5 & LPV & NO \\
HD206183 & O9.5IV-V & 8 & 29 & 33.7 ± 0.6 & 4.07 ± 0.08 & 3.45 ± 0.08 & 0.098 ± 0.017 & 5 & -13.5 & Q1.0 & 4 & SB1 & NO \\
HD193682 & O4.5IV(f) & 183 & 90 & 39.3 ± 1.3 & 3.72 ± 0.08 & 4.06 ± 0.09 & 0.12 ± 0.03 & 9 & -12.7 & Q1.7 & 5 & LS & YES \\
HD101190 & O6IV((f)) & 49 & 74 & 39.4 ± 0.9 & 3.85 ± 0.08 & 3.92 ± 0.08 & 0.088 ± 0.016 & 15 & -12.8 & Q1.5 & 3 & SB1 & NO \\
HD99897 & O6.5IV((f)) & 50 & 94 & 37.5 ± 0.6 & 3.56 ± 0.06 & 4.13 ± 0.06 & 0.111 ± 0.018 & 22 & -12.8 & Q1.0 & 1 & . & NO \\
HD101298 & O6.5IV((f)) & 71 & 69 & 37.9 ± 0.5 & 3.68 ± 0.05 & 4.02 ± 0.05 & 0.095 ± 0.020 & 16 & -12.7 & Q1.5 & 6 & LS & NO \\
HD322417 & O6.5IV((f)) & 68 & 100 & 38.0 ± 1.0 & 3.69 ± 0.13 & 4.02 ± 0.13 & 0.10 ± 0.03 & 19 & -12.8 & Q1.5 & 5 & SB1 & NO \\
HD124979 & O7.5IV(n)((f)) & 261 & 0 & 35.1 ± 1.0 & 3.60 ± 0.12 & 3.98 ± 0.13 & 0.091 ± 0.024 & 25 & -12.7 & Q2.3 & 6 & LS & YES \\
HD74920 & O7.5IVn((f)) & 291 & 0 & 35.8 ± 0.7 & 3.70 ± 0.06 & 3.91 ± 0.06 & 0.106 ± 0.018 & 30 & -12.8 & Q2.3 & 3 & LS & YES \\
HD135591 & O8IV((f)) & 60 & 60 & 35.1 ± 0.5 & 3.56 ± 0.08 & 4.01 ± 0.08 & 0.091 ± 0.016 & 9 & -13.1 & Q1.0 & 5 & LS & NO \\
HD41997 & O8IV(n)((f)) & 262 & 0 & 35.8 ± 0.8 & 3.66 ± 0.07 & 3.95 ± 0.08 & 0.106 ± 0.019 & 23 & -13.0 & Q1.8 & 8 & LS & YES \\
HD326331 & O8IVn((f)) & 332 & 0 & 34.8 ± 0.5 & 3.67 ± 0.05 & 3.89 ± 0.05 & 0.090 ± 0.017 & 30 & -12.7 & Q2.3 & 26 & LS & NO \\
HD46966 & O8.5IV & 40 & 66 & 35.8 ± 0.4 & 3.87 ± 0.06 & 3.74 ± 0.06 & 0.095 ± 0.015 & 9 & -13.1 & Q1.0 & 47 & LS & . \\
HD93028 & O9IV & 25 & 55 & 35.4 ± 0.9 & 3.99 ± 0.10 & 3.60 ± 0.11 & 0.09 ± 0.03 & 9 & -13.1 & Q1.0 & 7 & SB1 & NO \\
HD149452 & O9IVn & 318 & 0 & 33.7 ± 0.8 & 3.67 ± 0.08 & 3.82 ± 0.09 & 0.086 ± 0.025 & 19 & -13.1 & Q1.3 & 2 & . & NO \\
HD164438 & O9.2IV & 55 & 94 & 32.4 ± 0.8 & 3.51 ± 0.04 & 3.95 ± 0.04 & 0.09 ± 0.03 & 9 & -13.1 & Q1.0 & 9 & SB1 & YES \\
HD166852 & O9.2IV & 65 & 80 & 32.3 ± 1.2 & 3.41 ± 0.14 & 4.00 ± 0.15 & 0.09 ± 0.04 & 9 & -13.5 & Q1.0 & 1 & . & YES \\
HD57682 & O9.2IV & 7 & 34 & 35.1 ± 1.0 & 4.15 ± 0.14 & 3.42 ± 0.13 & 0.084 ± 0.016 & 7 & -13.2 & Q1.0 & 5 & LS & YES \\
HD76341 & O9.2IV & 51 & 93 & 33.1 ± 1.1 & 3.53 ± 0.08 & 3.90 ± 0.08 & 0.10 ± 0.04 & 7 & -13.2 & Q1.0 & 4 & LS & . \\
HD96622 & O9.2IV & 39 & 61 & 32.9 ± 0.8 & 3.64 ± 0.10 & 3.81 ± 0.11 & 0.102 ± 0.023 & 1 & -13.5 & Q1.0 & 4 & SB1 & NO \\
HD201345 & ON9.2IV & 79 & 89 & 33.6 ± 0.9 & 3.73 ± 0.12 & 3.77 ± 0.13 & 0.13 ± 0.04 & 9 & -13.5 & Q1.0 & 6 & LS & YES \\
HD166546 & O9.5IV & 38 & 73 & 32.5 ± 0.7 & 3.54 ± 0.07 & 3.89 ± 0.08 & 0.086 ± 0.016 & 11 & -13.5 & Q1.0 & 6 & LS & NO \\
HD192001 & O9.5IV & 44 & 57 & 33.4 ± 0.9 & 3.83 ± 0.09 & 3.63 ± 0.09 & 0.092 ± 0.022 & 1 & -13.5 & Q1.0 & 4 & SB1 & YES \\
HD202214 & O9.5IV & 26 & 36 & 32.2 ± 0.9 & 3.86 ± 0.15 & 3.56 ± 0.16 & 0.12 ± 0.03 & 1 & -13.5 & Q1.0 & 26 & SB1 & . \\
HD93027 & O9.5IV & 46 & 66 & 33.5 ± 0.8 & 3.89 ± 0.12 & 3.60 ± 0.12 & 0.11 ± 0.03 & 3 & -13.5 & Q1.0 & 2 & . & NO \\
HD164019 & O9.5IVp & 69 & 62 & 31.8 ± 0.5 & 3.41 ± 0.05 & 3.99 ± 0.05 & 0.078 ± 0.019 & 10 & -13.0 & Q1.0 & 3 & LS & NO \\
HD168941 & O9.5IVp & 83 & 101 & 31.8 ± 0.8 & 3.43 ± 0.08 & 3.98 ± 0.08 & 0.11 ± 0.04 & 10 & -13.3 & Q1.0 & 3 & LS & . \\
HD152199 & O9.5IV & 152 & 100 & 31.1 ± 0.9 & 3.57 ± 0.10 & 3.77 ± 0.11 & 0.11 ± 0.04 & 10 & -13.5 & Q1.2 & 1 & . & NO \\
HD36483 & O9.5IV(n) & 158 & 74 & 33.4 ± 1.0 & 3.82 ± 0.13 & 3.65 ± 0.14 & 0.09 ± 0.03 & 5 & -13.5 & Q1.2 & 6 & SB1 & NO \\
HD163892 & O9.5IV(n) & 215 & 0 & 32.8 ± 0.6 & 3.75 ± 0.08 & 3.70 ± 0.09 & 0.104 ± 0.021 & 8 & -13.5 & Q1.8 & 16 & SB1 & NO \\
HD161789 & O9.7IV & 28 & 20 & 33.3 ± 0.7 & 4.04 ± 0.09 & 3.43 ± 0.09 & 0.091 ± 0.016 & 7 & -13.5 & Q1.0 & 1 & . & NO \\
HD207538 & O9.7IV & 29 & 50 & 31.6 ± 0.6 & 3.81 ± 0.07 & 3.59 ± 0.07 & 0.108 ± 0.013 & 9 & -13.5 & Q1.0 & 6 & LS & NO \\
HD209339 & O9.7IV & 84 & 78 & 32.2 ± 0.7 & 3.80 ± 0.06 & 3.61 ± 0.07 & 0.12 ± 0.03 & 7 & -14.0 & Q1.0 & 5 & SB1 & NO \\
HD232525 & O9.7IV & 21 & 40 & 32.8 ± 1.1 & 3.86 ± 0.17 & 3.62 ± 0.18 & 0.11 ± 0.03 & 9 & -13.5 & Q1.0 & 1 & . & NO \\
HD152200 & O9.7IV(n) & 226 & 0 & 30.7 ± 0.6 & 3.72 ± 0.10 & 3.61 ± 0.11 & 0.11 ± 0.03 & 7 & -13.7 & Q1.3 & 21 & SB1 & NO \\
\noalign{\smallskip}
HD93250 & O4III(fc) & 70 & 84 & 44.7 ± 1.5 & 3.85 ± 0.10 & 4.14 ± 0.11 & 0.110 ± 0.024 & 30 & -12.5 & Q1.5 & 5 & LS & NO \\
HD93843 & O5III(fc) & 58 & 120 & 37.5 ± 0.8 & 3.49 ± 0.03 & 4.21 ± 0.03 & 0.12 ± 0.03 & 12 & -12.5 & Q3.0 & 5 & LS & NO \\
HD97253 & O5III(f) & 70 & 105 & 39.4 ± 0.8 & 3.62 ± 0.05 & 4.14 ± 0.05 & 0.107 ± 0.016 & 30 & -12.3 & Q3.0 & 4 & LS & NO \\
HD96946 & O6.5III(f) & 72 & 58 & 38.6 ± 0.9 & 3.84 ± 0.10 & 3.90 ± 0.11 & 0.11 ± 0.03 & 9 & -12.7 & Q1.5 & 5 & SB1 & NO \\
BD+60261 & O7.5III(n)((f)) & 162 & 92 & 35.0 ± 0.8 & 3.53 ± 0.05 & 4.03 ± 0.06 & 0.099 ± 0.025 & 22 & -12.7 & Q2.2 & 3 & LS & YES \\
HD203064 & O7.5IIIn((f)) & 312 & 0 & 35.5 ± 0.7 & 3.78 ± 0.07 & 3.82 ± 0.07 & 0.095 ± 0.018 & 20 & -12.9 & Q2.3 & 32 & LS & . \\
HD24912 & O7.5III(n)((f)) & 230 & 0 & 35.7 ± 0.7 & 3.66 ± 0.05 & 3.96 ± 0.06 & 0.107 ± 0.014 & 25 & -12.7 & Q2.3 & 108 & LS & . \\
HD97434 & O7.5III(n)((f)) & 217 & 0 & 34.7 ± 0.5 & 3.56 ± 0.04 & 4.00 ± 0.04 & 0.106 ± 0.022 & 15 & -12.7 & Q2.3 & 3 & LS & NO \\
HD36861 & O8III((f)) & 52 & 77 & 35.5 ± 0.6 & 3.54 ± 0.06 & 4.04 ± 0.06 & 0.096 ± 0.014 & 15 & -13.1 & Q1.0 & 963 & LS & . \\
HD173820 & O8III & 65 & 75 & 34.9 ± 1.4 & 3.42 ± 0.12 & 4.10 ± 0.14 & 0.11 ± 0.04 & 20 & -13.0 & Q1.0 & 1 & . & NO \\
HD96638 & O8III & 249 & 0 & 33.8 ± 1.0 & 3.50 ± 0.08 & 4.00 ± 0.09 & 0.079 ± 0.018 & 30 & -12.8 & Q2.3 & 1 & . & YES \\
HD218195 & O8.5IIINstr & 44 & 78 & 34.5 ± 0.7 & 3.60 ± 0.05 & 3.96 ± 0.05 & 0.116 ± 0.018 & 14 & -13.1 & Q1.0 & 6 & LPV & NO \\
HD13268 & ON8.5IIIn & 290 & 0 & 34.7 ± 0.7 & 3.72 ± 0.09 & 3.83 ± 0.09 & 0.113 ± 0.018 & 25 & -13.0 & Q1.3 & 9 & LS & NO \\
HD116852 & O8.5II-III((f)) & 114 & 69 & 33.1 ± 0.8 & 3.32 ± 0.05 & 4.13 ± 0.05 & 0.081 ± 0.019 & 18 & -12.7 & Q1.7 & 3 & LS & YES \\
HD305523 & O9II-III & 57 & 78 & 32.3 ± 0.7 & 3.41 ± 0.05 & 4.00 ± 0.05 & 0.11 ± 0.03 & 1 & -13.0 & Q1.0 & 1 & . & NO \\
HD109978 & O9III & 53 & 74 & 32.8 ± 1.1 & 3.37 ± 0.16 & 4.10 ± 0.17 & 0.11 ± 0.03 & 30 & -12.7 & Q1.0 & 2 & . & NO \\
HD113904B & O9III & 84 & 83 & 32.9 ± 0.4 & 3.52 ± 0.02 & 3.94 ± 0.02 & 0.095 ± 0.015 & 8 & -13.1 & Q1.0 & 5 & SB1 & NO \\
HD60369 & O9III & 78 & 45 & 32.7 ± 0.7 & 3.45 ± 0.07 & 3.98 ± 0.08 & 0.079 ± 0.021 & 9 & -13.2 & Q1.0 & 4 & SB1 & NO \\
HD96654 & O9III & 113 & 65 & 33.2 ± 0.8 & 3.57 ± 0.07 & 3.91 ± 0.07 & 0.087 ± 0.021 & 5 & -13.5 & Q1.2 & 2 & . & NO \\
HD24431 & O9III & 49 & 81 & 34.5 ± 0.7 & 3.70 ± 0.05 & 3.86 ± 0.05 & 0.079 ± 0.021 & 5 & -13.5 & Q1.5 & 51 & SB1 & . \\
HD16832 & O9.2III & 45 & 70 & 32.2 ± 0.8 & 3.31 ± 0.05 & 4.11 ± 0.05 & 0.085 ± 0.022 & 16 & -13.0 & Q1.0 & 3 & LS & NO \\
CPD-352105 & O9.2III & 79 & 59 & 32.1 ± 0.5 & 3.47 ± 0.09 & 3.91 ± 0.10 & 0.079 ± 0.019 & 1 & -13.1 & Q1.0 & 1 & . & . \\
HD90087 & O9.2III(n) & 293 & 0 & 32.5 ± 0.8 & 3.59 ± 0.06 & 3.83 ± 0.08 & 0.12 ± 0.04 & 20 & -12.9 & Q2.3 & 5 & LS & NO \\
HD152247 & O9.2III & 82 & 96 & 32.2 ± 1.4 & 3.39 ± 0.10 & 3.97 ± 0.11 & 0.10 ± 0.05 & 1 & -12.9 & Q3.0 & 7 & SB1 & NO \\
HD52266 & O9.5IIIn & 267 & 0 & 32.2 ± 0.8 & 3.62 ± 0.11 & 3.79 ± 0.12 & 0.12 ± 0.04 & 18 & -13.0 & Q2.3 & 11 & LS & NO \\
HD93521 & O9.5IIInn & 385 & 0 & 31.9 ± 0.7 & 3.77 ± 0.07 & 3.64 ± 0.07 & 0.12 ± 0.04 & 25 & -12.9 & Q2.3 & 94 & LS & NO \\
HD112784 & O9.7III & 36 & 51 & 31.5 ± 0.7 & 3.59 ± 0.10 & 3.78 ± 0.10 & 0.090 ± 0.022 & 12 & -13.5 & Q1.0 & 2 & . & NO \\
HD118198 & O9.7III & 43 & 56 & 31.5 ± 0.7 & 3.48 ± 0.09 & 3.91 ± 0.10 & 0.10 ± 0.04 & 9 & -13.1 & Q1.0 & 2 & . & NO \\
HD150475 & O9.7III & 86 & 60 & 32.1 ± 0.6 & 3.64 ± 0.09 & 3.76 ± 0.10 & 0.090 ± 0.015 & 9 & -13.7 & Q1.0 & 1 & . & YES \\
HD156234 & O9.7III & 90 & 70 & 30.3 ± 0.8 & 3.40 ± 0.12 & 3.91 ± 0.13 & 0.082 ± 0.025 & 10 & -13.5 & Q1.0 & 3 & SB1 & NO \\
HD55879 & O9.7III & 28 & 61 & 31.5 ± 0.7 & 3.54 ± 0.09 & 3.84 ± 0.10 & 0.11 ± 0.03 & 10 & -13.5 & Q1.0 & 9 & LS & NO \\
HD152622 & O9.7III & 159 & 43 & 30.5 ± 0.8 & 3.53 ± 0.10 & 3.79 ± 0.11 & 0.12 ± 0.04 & 9 & -13.5 & Q1.2 & 1 & . & NO \\
HD168183 & O9.7III & 110 & 54 & 30.3 ± 1.4 & 3.53 ± 0.16 & 3.76 ± 0.17 & 0.09 ± 0.04 & 1 & -13.5 & Q1.2 & 3 & SB1 & . \\
HD13022 & O9.7III & 110 & 65 & 30.1 ± 0.7 & 3.23 ± 0.05 & 4.07 ± 0.05 & 0.11 ± 0.04 & 12 & -13.0 & Q1.2 & 3 & LS & NO \\
HD154643 & O9.7III & 101 & 78 & 31.3 ± 0.8 & 3.51 ± 0.11 & 3.86 ± 0.12 & 0.10 ± 0.04 & 15 & -13.4 & Q1.7 & 5 & SB1 & NO \\
HD306099 & O9.7III & 154 & 85 & 30.7 ± 0.8 & 3.63 ± 0.08 & 3.71 ± 0.09 & 0.078 ± 0.020 & 9 & -12.7 & Q3.2 & 1 & . & NO \\
HD37737 & O9.5II-III(n) & 201 & 0 & 30.7 ± 0.8 & 3.60 ± 0.08 & 3.76 ± 0.08 & 0.09 ± 0.03 & 11 & -13.5 & Q1.3 & 19 & SB1 & . \\
HD15137 & O9.5II-IIIn & 270 & 0 & 30.6 ± 0.9 & 3.44 ± 0.07 & 3.87 ± 0.08 & 0.13 ± 0.05 & 16 & -13.1 & Q1.8 & 18 & SB1 & YES \\
\noalign{\smallskip}
HD152233 & O6II(f) & 62 & 105 & 37.4 ± 0.8 & 3.64 ± 0.07 & 4.04 ± 0.08 & 0.12 ± 0.04 & 15 & -12.5 & Q3.0 & 54 & SB1 & NO \\
HD157857 & O6.5II(f) & 114 & 69 & 36.3 ± 0.8 & 3.57 ± 0.08 & 4.07 ± 0.09 & 0.116 ± 0.022 & 17 & -12.5 & Q2.2 & 9 & LS & YES \\
HD151515 & O7II(f) & 67 & 98 & 35.9 ± 1.1 & 3.53 ± 0.12 & 4.08 ± 0.13 & 0.11 ± 0.04 & 18 & -12.7 & Q1.0 & 5 & LS & NO \\
HD167659 & O7II-III(f) & 71 & 84 & 36.7 ± 0.7 & 3.60 ± 0.05 & 4.06 ± 0.05 & 0.092 ± 0.014 & 15 & -12.7 & Q1.5 & 5 & LS & NO \\
BD-134927 & O7II(f) & 98 & 78 & 35.7 ± 0.8 & 3.48 ± 0.08 & 4.11 ± 0.08 & 0.093 ± 0.018 & 25 & -12.6 & Q3.0 & 6 & LS & NO \\
HD94963 & O7II(f) & 76 & 92 & 35.8 ± 0.8 & 3.50 ± 0.05 & 4.11 ± 0.05 & 0.089 ± 0.018 & 20 & -12.6 & Q3.0 & 3 & LS & NO \\
HD35633 & O7.5II(n)(f)p & 170 & 50 & 33.6 ± 1.2 & 3.33 ± 0.08 & 4.17 ± 0.09 & 0.081 ± 0.018 & 30 & -12.6 & Q2.2 & 6 & SB1 & NO \\
HD194334 & O7.5II-III & 61 & 89 & 35.0 ± 0.5 & 3.41 ± 0.05 & 4.16 ± 0.05 & 0.098 ± 0.010 & 21 & -12.7 & Q3.0 & 4 & LS & NO \\
HD162978 & O8II((f)) & 54 & 86 & 34.9 ± 0.6 & 3.51 ± 0.03 & 4.05 ± 0.03 & 0.085 ± 0.012 & 21 & -12.7 & Q1.5 & 30 & LS & . \\
HD74194 & O8.5Ib-II(f)p & 184 & 49 & 32.5 ± 0.7 & 3.28 ± 0.04 & 4.14 ± 0.05 & 0.101 ± 0.016 & 24 & -12.5 & Q2.2 & 21 & SB1 & YES \\
HD75211 & O8.5II((f)) & 145 & 53 & 33.4 ± 0.7 & 3.35 ± 0.04 & 4.12 ± 0.05 & 0.111 ± 0.022 & 25 & -12.6 & Q2.2 & 5 & SB1 & YES \\
HD71304 & O9II & 62 & 84 & 32.2 ± 0.5 & 3.19 ± 0.11 & 4.18 ± 0.12 & 0.11 ± 0.03 & 12 & -12.7 & Q1.0 & 2 & . & NO \\
HD57061 & O9II & 57 & 85 & 32.9 ± 1.1 & 3.42 ± 0.12 & 4.02 ± 0.12 & 0.075 ± 0.022 & 1 & -12.7 & Q3.0 & 11 & SB1 & . \\
HD36486 & O9.5IINwk & 100 & 94 & 30.7 ± 1.1 & 3.34 ± 0.05 & 3.97 ± 0.06 & 0.07 ± 0.03 & 3 & -13.0 & Q1.7 & 149 & SB1 & . \\
HD305619 & O9.7II & 48 & 94 & 31.2 ± 0.5 & 3.28 ± 0.05 & 4.08 ± 0.05 & 0.092 ± 0.017 & 22 & -12.7 & Q1.0 & 1 & . & NO \\
HD152405 & O9.7II & 59 & 61 & 30.5 ± 0.7 & 3.32 ± 0.08 & 4.00 ± 0.09 & 0.11 ± 0.04 & 5 & -13.0 & Q1.0 & 8 & SB1 & NO \\
HD68450 & O9.7II & 39 & 77 & 30.5 ± 0.7 & 3.30 ± 0.06 & 4.05 ± 0.06 & 0.10 ± 0.03 & 11 & -13.1 & Q1.0 & 4 & LS & NO \\
HD28446A & O9.7IIn & 299 & 0 & 29.4 ± 0.8 & 3.66 ± 0.07 & 3.62 ± 0.08 & 0.086 ± 0.025 & 16 & -13.5 & Q1.3 & 9 & SB1 & NO \\
HD167411 & O9.7II & 85 & 89 & 28.8 ± 1.2 & 3.14 ± 0.10 & 4.04 ± 0.12 & 0.09 ± 0.04 & 9 & -13.0 & Q1.5 & 3 & LS & NO \\
HD10125 & O9.7II & 122 & 125 & 30.7 ± 0.8 & 3.33 ± 0.05 & 4.02 ± 0.05 & 0.12 ± 0.04 & 15 & -12.7 & Q2.2 & 5 & SB1 & . \\
HD13745 & O9.7II(n) & 197 & 57 & 30.3 ± 0.7 & 3.23 ± 0.05 & 4.08 ± 0.05 & 0.098 ± 0.021 & 30 & -12.8 & Q2.2 & 5 & LS & YES \\
HD165174 & O9.7IIn & 264 & 0 & 30.7 ± 1.2 & 3.49 ± 0.08 & 3.82 ± 0.09 & 0.12 ± 0.04 & 20 & -13.0 & Q2.3 & 11 & LS & NO \\
HD69106 & O9.7IIn & 277 & 0 & 30.3 ± 0.9 & 3.58 ± 0.05 & 3.67 ± 0.06 & 0.10 ± 0.04 & 1 & -13.4 & Q3.3 & 3 & LPV & NO \\
HD15570 & O4If & 81 & 115 & 40.9 ± 1.6 & 3.66 ± 0.09 & 4.18 ± 0.10 & 0.104 ± 0.022 & 30 & -11.9 & Q1.5 & 8 & LS & NO \\
HD193514 & O7Ib(f) & 73 & 84 & 35.6 ± 0.8 & 3.54 ± 0.10 & 4.06 ± 0.10 & 0.084 ± 0.024 & 20 & -12.5 & Q3.0 & 4 & LS & NO \\
HD96917 & O8Ib(n)(f) & 165 & 77 & 32.0 ± 0.6 & 3.25 ± 0.05 & 4.16 ± 0.05 & 0.117 ± 0.015 & 30 & -12.4 & Q2.2 & 4 & LS & YES \\
HD112244 & O8.5Iab(f)p & 124 & 80 & 31.3 ± 0.8 & 3.19 ± 0.10 & 4.17 ± 0.11 & 0.11 ± 0.03 & 25 & -12.3 & Q2.2 & 12 & SB1 & . \\
BD+391328 & O8.5Iab(n)(f) & 90 & 37 & 33.7 ± 1.4 & 3.37 ± 0.19 & 4.09 ± 0.20 & 0.079 ± 0.021 & 17 & -12.5 & Q3.0 & 1 & . & YES \\
HD292167 & OC8.5Ib & 112 & 73 & 31.1 ± 1.5 & 3.18 ± 0.17 & 4.18 ± 0.18 & 0.09 ± 0.04 & 18 & -12.5 & Q3.2 & 1 & . & . \\
HD210809 & O9Iab & 70 & 105 & 31.0 ± 0.5 & 3.09 ± 0.05 & 4.25 ± 0.05 & 0.13 ± 0.03 & 30 & -12.5 & Q2.0 & 5 & LS & . \\
HD30614 & O9Ia & 113 & 77 & 30.0 ± 1.0 & 3.01 ± 0.05 & 4.27 ± 0.06 & 0.11 ± 0.03 & 25 & -12.4 & Q2.2 & 212 & SB1 & . \\
HD202124 & O9Iab & 88 & 114 & 30.7 ± 0.8 & 3.05 ± 0.05 & 4.26 ± 0.06 & 0.12 ± 0.04 & 25 & -12.5 & Q3.0 & 4 & LS & YES \\
HD237211 & O9Ib & 51 & 82 & 30.5 ± 0.7 & 3.17 ± 0.05 & 4.17 ± 0.05 & 0.12 ± 0.03 & 25 & -12.6 & Q3.0 & 4 & LS & NO \\
HD152249 & OC9Iab & 71 & 70 & 31.5 ± 0.7 & 3.27 ± 0.05 & 4.12 ± 0.06 & 0.082 ± 0.022 & 20 & -12.5 & Q3.0 & 21 & LS & NO \\
CPD-595634 & O9.2Ib & 72 & 65 & 31.8 ± 0.5 & 3.30 ± 0.05 & 4.10 ± 0.05 & 0.095 ± 0.022 & 11 & -12.7 & Q1.0 & 2 & . & NO \\
HD60479 & O9.2Ib & 55 & 82 & 31.1 ± 0.8 & 3.25 ± 0.08 & 4.11 ± 0.09 & 0.083 ± 0.023 & 25 & -12.9 & Q1.5 & 2 & . & NO \\
HD154368 & O9.2Iab & 65 & 78 & 30.6 ± 0.7 & 3.10 ± 0.10 & 4.22 ± 0.11 & 0.13 ± 0.04 & 25 & -12.5 & Q3.0 & 5 & LS & NO \\
HD76968 & O9.2Ib & 53 & 65 & 31.1 ± 0.7 & 3.26 ± 0.10 & 4.10 ± 0.10 & 0.09 ± 0.03 & 13 & -12.5 & Q3.0 & 6 & SB1 & YES \\
HD167330 & O9.5Iab & 83 & 106 & 30.6 ± 0.8 & 3.15 ± 0.10 & 4.19 ± 0.11 & 0.11 ± 0.03 & 25 & -12.5 & Q3.0 & 4 & LS & NO \\
HD18409 & O9.7Ib & 131 & 102 & 30.2 ± 0.5 & 3.15 ± 0.04 & 4.16 ± 0.05 & 0.095 ± 0.018 & 21 & -12.7 & Q2.2 & 4 & LS & YES \\
HD167264 & O9.7Iab & 71 & 69 & 29.0 ± 0.5 & 3.11 ± 0.10 & 4.10 ± 0.11 & 0.084 ± 0.022 & 19 & -12.7 & Q3.0 & 10 & SB1 & . \\
HD47432 & O9.7Ib & 97 & 63 & 29.2 ± 0.5 & 3.07 ± 0.06 & 4.20 ± 0.07 & 0.097 ± 0.020 & 21 & -12.5 & Q3.0 & 5 & LS & NO \\
HD152147 & O9.7IbNwk & 91 & 64 & 29.6 ± 0.7 & 3.16 ± 0.07 & 4.10 ± 0.07 & 0.09 ± 0.03 & 7 & -12.7 & Q3.0 & 4 & SB1 & NO \\
HD104565 & OC9.7Ia & 56 & 116 & 28.9 ± 1.2 & 2.93 ± 0.05 & 4.26 ± 0.06 & 0.09 ± 0.03 & 30 & -12.4 & Q3.0 & 1 & . & YES \\
HD154811 & OC9.7Ib & 124 & 62 & 29.6 ± 0.6 & 3.20 ± 0.06 & 4.08 ± 0.07 & 0.09 ± 0.03 & 11 & -12.7 & Q3.2 & 2 & . & NO \\
BD+364063 & ON9.7Ib & 117 & 49 & 27.4 ± 1.0 & 3.04 ± 0.13 & 4.11 ± 0.15 & 0.09 ± 0.04 & 3 & -12.3 & Q3.2 & 6 & SB1 & NO \\
HD149038 & O9.7Iab & 52 & 90 & 29.6 ± 0.6 & 3.15 ± 0.07 & 4.12 ± 0.08 & 0.081 ± 0.020 & 17 & -12.7 & Q3.0 & 28 & LS & . \\
HD94370 & O7(n)fp & 182 & 91 & 35.2 ± 0.4 & 3.47 ± 0.04 & 4.11 ± 0.05 & 0.089 ± 0.014 & 19 & -12.5 & Q2.2 & 8 & LS & NO \\
\noalign{\smallskip}
     \end{longtable}
}

	\onecolumn
	\pagestyle{empty}
	\fontsize{8}{8}\selectfont
	\onecolumn
	{
		\noindent\begin{longtable}{l@{\hskip 0.1in}l@{\hskip 0.1in}r@{\hskip 0.1in}r@{\hskip 0.1in}c@{\hskip 0.1in}c@{\hskip 0.1in}c@{\hskip 0.1in}c@{\hskip 0.1in}c@{\hskip 0.1in}c@{\hskip 0.1in}c@{\hskip 0.1in}c@{\hskip 0.1in}c@{\hskip 0.1in}c}
			\captionsetup{justification=raggedright, singlelinecheck=false}
			\caption{Spectroscopic parameters for the sample analyzed in this work and identified as He-rich. }\\
			\multicolumn{14}{l}{\parbox{16.7cm}{Same columns as Table~\ref{tableResults_low}}}\\
			\label{tableResults_high}\\
			\hline \hline
			\noalign{\smallskip}
			STAR-ID & SpC & $v \sin i$ & \vmacro & \Teff & \gravc & \Lsp & \helio & \micro & logQ  & Qual. & \# sp & SB     & RW \\
			&     & [\kms]    & [\kms] &  [kK] & [dex] & [dex]  & [dex] & [\kms] & [dex] & flag  &     & status & status \\
			\hline
			\noalign{\smallskip}
			\endfirsthead
			\caption{continued.}\\
			\hline
			\hline
			\noalign{\smallskip}
			STAR-ID & SpC & $v \sin i$ & \vmacro & \Teff & \gravc & \Lsp & \helio & \micro & logQ  & Qual. & \# sp & SB     & RW \\
			&     & [\kms]    & [\kms] &  [kK] & [dex] & [dex]  & [dex] & [\kms] & [dex] & flag  &     & status & status \\
			\hline
			\noalign{\smallskip}
			\endhead
			\noalign{\smallskip}
			\noalign{\smallskip}
			\hline\hline
			\noalign{\smallskip}
			\multicolumn{14}{l}{SB status: \textbf{LS}: Likely single, \textbf{LPV}: Line profile variable, \textbf{SB1}: Single-lined spectroscopic binary.}\\
			\multicolumn{14}{l}{Classifications based on fewer than three spectra (as indicated in the \# sp. column) should be treated with caution.}
			\endfoot
HD256725 & O5V((fc))z & 67 & 38 & 41.3 ± 1.4 & 3.95 ± 0.14 & 3.91 ± 0.15 & 0.15 ± 0.05 & 9 & -13.1 & Q1.0 & 4 & LS & NO \\
BD+60134 & O5.5V(n)((f)) & 250 & 0 & 41.2 ± 1.7 & 4.03 ± 0.18 & 3.83 ± 0.20 & >0.20 ± 0.07 & 18 & -13.0 & Q1.3 & 4 & LS & YES \\
BD+602635 & O6V((f)) & 51 & 73 & 39.7 ± 1.0 & 3.78 ± 0.10 & 4.01 ± 0.10 & 0.20 ± 0.05 & 19 & -13.1 & Q1.0 & 5 & LS & NO \\
ALS5039 & ON6V((f))z & 124 & 0 & 41.5 ± 1.5 & 4.22 ± 0.04 & 3.63 ± 0.05 & 0.17 ± 0.06 & 16 & -13.1 & Q1.2 & 7 & LS & YES \\
HD12993 & ON6.5V((f)) & 84 & 79 & 39.2 ± 1.0 & 3.88 ± 0.14 & 3.88 ± 0.15 & 0.17 ± 0.05 & 14 & -13.1 & Q1.0 & 3 & LS & YES \\
HD228841 & O6.5Vn((f)) & 317 & 0 & 37.8 ± 1.2 & 3.89 ± 0.10 & 3.82 ± 0.11 & 0.15 ± 0.06 & 16 & -13.3 & Q1.3 & 10 & LS & YES \\
HD167633 & O6.5V((f)) & 129 & 106 & 37.6 ± 0.8 & 3.68 ± 0.10 & 4.02 ± 0.10 & 0.17 ± 0.05 & 9 & -13.0 & Q1.7 & 5 & LS & NO \\
BD+56594 & O7Vz & 27 & 38 & 38.3 ± 0.8 & 3.98 ± 0.12 & 3.74 ± 0.12 & >0.14 ± 0.03 & 10 & -13.5 & Q1.0 & 3 & LS & NO \\
HD193595 & ON7V((f)) & 41 & 61 & 37.7 ± 0.8 & 3.70 ± 0.10 & 3.99 ± 0.10 & 0.15 ± 0.03 & 20 & -13.0 & Q1.0 & 5 & LS & NO \\
HD90273 & ON7V((f)) & 55 & 55 & 38.2 ± 0.7 & 3.72 ± 0.07 & 4.00 ± 0.07 & 0.18 ± 0.04 & 9 & -13.0 & Q1.0 & 1 & . & NO \\
HD110360 & ON7Vz & 96 & 86 & 38.8 ± 1.3 & 4.08 ± 0.15 & 3.69 ± 0.16 & 0.18 ± 0.05 & 9 & -13.1 & Q1.0 & 2 & SB1 & NO \\
HD5689 & O7Vn((f)) & 255 & 0 & 37.0 ± 1.1 & 3.68 ± 0.11 & 3.98 ± 0.13 & 0.17 ± 0.05 & 20 & -13.1 & Q1.3 & 6 & LS & YES \\
HD41161 & O8Vn & 331 & 0 & 35.7 ± 0.7 & 3.83 ± 0.06 & 3.78 ± 0.07 & 0.13 ± 0.03 & 25 & -13.3 & Q1.3 & 8 & LS & NO \\
HD14633 & ON8.5V & 121 & 0 & 35.1 ± 0.5 & 3.73 ± 0.12 & 3.81 ± 0.13 & 0.17 ± 0.04 & 9 & -13.5 & Q1.2 & 12 & SB1 & YES \\
HD48279 & ON8.5Vz & 131 & 74 & 36.1 ± 0.9 & 3.88 ± 0.11 & 3.74 ± 0.12 & 0.15 ± 0.03 & 13 & -13.5 & Q1.2 & 4 & LS & NO \\
HD12323 & ON9.2V & 121 & 82 & 34.3 ± 0.9 & 3.96 ± 0.18 & 3.58 ± 0.19 & 0.18 ± 0.05 & 9 & -14.0 & Q1.2 & 19 & SB1 & YES \\
HD36512 & O9.7V & 13 & 33 & 33.3 ± 0.7 & 4.11 ± 0.13 & 3.38 ± 0.13 & 0.121 ± 0.022 & 5 & -14.0 & Q1.0 & 145 & LS & NO \\
CPD-417721A & O9.7V:(n) & 191 & 66 & 31.6 ± 0.7 & 3.77 ± 0.04 & 3.60 ± 0.05 & 0.14 ± 0.04 & 5 & -13.5 & Q1.2 & 3 & LS & NO \\
HD192281 & O4.5IV(n)(f) & 277 & 0 & 41.0 ± 1.3 & 3.87 ± 0.08 & 3.99 ± 0.09 & >0.25 ± 0.07 & 10 & -12.8 & Q2.3 & 9 & LS & YES \\
HD63005 & O6.5IV & 56 & 66 & 38.4 ± 1.7 & 3.75 ± 0.18 & 3.96 ± 0.19 & 0.13 ± 0.05 & 19 & -13.1 & Q1.0 & 2 & . & NO \\
HD94024 & O8IV & 162 & 52 & 34.6 ± 0.6 & 3.55 ± 0.04 & 4.01 ± 0.04 & 0.14 ± 0.04 & 15 & -13.2 & Q1.2 & 8 & SB1 & YES \\
HD102415 & ON9IV:nn & 366 & 0 & 33.5 ± 1.1 & 3.94 ± 0.10 & 3.54 ± 0.11 & >0.20 ± 0.03 & 9 & -13.5 & Q1.8 & 6 & SB1 & NO \\
HD149757 & O9.2IVnn & 385 & 0 & 31.7 ± 0.7 & 3.83 ± 0.06 & 3.58 ± 0.06 & 0.14 ± 0.03 & 9 & -13.0 & Q1.8 & 214 & LS & . \\
HD119547 & O9.5IV(n) & 183 & 0 & 34.1 ± 1.4 & 4.01 ± 0.30 & 3.50 ± 0.31 & 0.14 ± 0.06 & 9 & -13.5 & Q1.2 & 1 & . & NO \\
HD123056 & O9.5IV(n) & 193 & 26 & 31.6 ± 0.6 & 3.63 ± 0.06 & 3.74 ± 0.07 & 0.15 ± 0.04 & 9 & -13.2 & Q1.2 & 9 & SB1 & . \\
HD308813 & O9.7IV(n) & 215 & 0 & 31.7 ± 0.7 & 3.91 ± 0.07 & 3.49 ± 0.07 & 0.13 ± 0.03 & 7 & -13.5 & Q1.3 & 5 & SB1 & NO \\
\noalign{\smallskip}
HD338931 & O6III(f) & 170 & 0 & 38.7 ± 1.4 & 4.02 ± 0.09 & 3.77 ± 0.10 & 0.14 ± 0.04 & 25 & -12.5 & Q1.2 & 3 & LS & YES \\
HD190864 & ON6.5III(f) & 66 & 90 & 37.4 ± 0.8 & 3.56 ± 0.07 & 4.13 ± 0.08 & 0.144 ± 0.023 & 23 & -12.8 & Q1.0 & 4 & LS & NO \\
HD130298 & O6.5III(n)(f) & 167 & 85 & 37.4 ± 0.8 & 3.61 ± 0.07 & 4.08 ± 0.08 & 0.16 ± 0.03 & 22 & -12.7 & Q1.7 & 10 & SB1 & YES \\
HD186980 & O7.5III((f)) & 61 & 83 & 35.5 ± 0.6 & 3.44 ± 0.05 & 4.14 ± 0.05 & 0.13 ± 0.03 & 20 & -12.9 & Q1.0 & 4 & LS & YES \\
HD105627 & O9III & 141 & 122 & 33.2 ± 0.6 & 3.48 ± 0.08 & 3.99 ± 0.08 & 0.14 ± 0.04 & 11 & -13.1 & Q1.7 & 5 & SB1 & YES \\
HD191423 & ON9II-IIInn & 432 & 0 & 32.5 ± 0.9 & 3.76 ± 0.08 & 3.69 ± 0.09 & 0.14 ± 0.06 & 25 & -13.1 & Q2.3 & 10 & LS & YES \\
HD150574 & ON9III(n) & 252 & 0 & 33.1 ± 1.0 & 3.64 ± 0.12 & 3.85 ± 0.12 & 0.14 ± 0.04 & 20 & -13.0 & Q2.3 & 1 & . & NO \\
HD117490 & ON9.5IIInn & 343 & 0 & 32.3 ± 0.8 & 3.81 ± 0.06 & 3.59 ± 0.07 & 0.14 ± 0.04 & 13 & -13.2 & Q1.8 & 10 & LS & YES \\
HD91651 & ON9.5IIIn & 271 & 0 & 32.8 ± 0.6 & 3.64 ± 0.06 & 3.71 ± 0.08 & 0.13 ± 0.04 & 13 & -12.9 & Q2.3 & 9 & LS & NO \\
HD15642 & O9.5II-IIIn & 286 & 0 & 29.9 ± 0.9 & 3.52 ± 0.08 & 3.77 ± 0.09 & 0.18 ± 0.10 & 15 & -13.1 & Q2.3 & 10 & LS & YES \\
\noalign{\smallskip}
HD228368 & O7II & 267 & 0 & 34.3 ± 1.4 & 3.46 ± 0.10 & 4.06 ± 0.12 & 0.13 ± 0.05 & 30 & -12.6 & Q2.3 & 1 & . & YES \\
HD34656 & O7.5II(f) & 67 & 77 & 35.9 ± 0.5 & 3.51 ± 0.03 & 4.11 ± 0.03 & 0.16 ± 0.03 & 21 & -12.7 & Q1.0 & 111 & LS & . \\
HD171589 & O7.5II(f) & 100 & 86 & 36.2 ± 1.0 & 3.63 ± 0.13 & 4.02 ± 0.14 & 0.130 ± 0.025 & 17 & -12.7 & Q1.2 & 6 & LS & YES \\
HD175754 & O8II(n)((f))p & 182 & 121 & 33.6 ± 0.8 & 3.34 ± 0.05 & 4.16 ± 0.05 & 0.15 ± 0.03 & 30 & -12.6 & Q2.2 & 7 & LS & YES \\
HD207198 & O8.5II((f)) & 52 & 97 & 33.3 ± 0.7 & 3.33 ± 0.05 & 4.14 ± 0.05 & 0.13 ± 0.03 & 20 & -12.8 & Q1.5 & 70 & LS & NO \\
HD89137 & ON9.7II(n) & 238 & 0 & 29.5 ± 0.8 & 3.44 ± 0.09 & 3.83 ± 0.10 & 0.18 ± 0.05 & 16 & -13.0 & Q2.3 & 4 & LS & YES \\
HD190429A & O4If & 90 & 113 & 40.9 ± 1.7 & 3.69 ± 0.12 & 4.15 ± 0.14 & 0.21 ± 0.07 & 30 & -11.9 & Q1.5 & 4 & LS & NO \\
HD14947 & O4.5If & 114 & 22 & 37.8 ± 1.0 & 3.59 ± 0.09 & 4.10 ± 0.10 & 0.125 ± 0.022 & 7 & -12.1 & Q1.7 & 4 & LPV & . \\
HD169582 & O6Iaf & 66 & 97 & 38.4 ± 1.8 & 3.67 ± 0.22 & 4.06 ± 0.23 & >0.25 ± 0.10 & 25 & -12.3 & Q3.0 & 4 & LS & NO \\
HD172175 & O6.5I(n)fp & 237 & 0 & 37.3 ± 0.9 & 3.60 ± 0.06 & 4.07 ± 0.07 & 0.16 ± 0.04 & 25 & -12.4 & Q2.3 & 5 & LS & YES \\
HD210839 & O6.5Iab:(n)fp & 214 & 0 & 36.3 ± 0.7 & 3.57 ± 0.08 & 4.05 ± 0.08 & 0.15 ± 0.04 & 30 & -12.3 & Q2.3 & 103 & LS & . \\
CPD-262716 & O6.5Iabf & 139 & 138 & 35.8 ± 1.0 & 3.53 ± 0.06 & 4.09 ± 0.07 & 0.15 ± 0.04 & 30 & -12.3 & Q3.2 & 5 & LS & YES \\
HD195213 & O7Ib(f) & 62 & 85 & 35.1 ± 1.0 & 3.37 ± 0.11 & 4.20 ± 0.11 & 0.19 ± 0.07 & 25 & -12.5 & Q3.0 & 3 & LS & NO \\
HD69464 & O7Ib(f) & 73 & 105 & 35.8 ± 1.2 & 3.38 ± 0.09 & 4.22 ± 0.11 & 0.14 ± 0.04 & 25 & -12.5 & Q3.0 & 8 & LS & NO \\
HD156154 & O7.5Ib(f) & 62 & 102 & 34.6 ± 0.8 & 3.39 ± 0.06 & 4.17 ± 0.06 & 0.128 ± 0.021 & 25 & -12.5 & Q2.0 & 4 & LS & NO \\
HD17603 & O7.5Ib(f) & 103 & 108 & 33.3 ± 0.9 & 3.25 ± 0.07 & 4.23 ± 0.08 & 0.16 ± 0.05 & 25 & -12.4 & Q2.2 & 4 & LS & NO \\
HD120521 & O7.5Ib(f) & 100 & 95 & 33.3 ± 1.0 & 3.26 ± 0.07 & 4.22 ± 0.08 & 0.14 ± 0.04 & 25 & -12.5 & Q2.2 & 2 & LPV/ & YES \\
HD192639 & O7.5Iabf & 82 & 95 & 34.3 ± 0.8 & 3.32 ± 0.05 & 4.20 ± 0.05 & 0.14 ± 0.04 & 30 & -12.3 & Q3.0 & 13 & . & YES \\
HD188001 & O7.5Iabf & 69 & 100 & 33.5 ± 0.9 & 3.13 ± 0.05 & 4.29 ± 0.05 & 0.16 ± 0.06 & 25 & -12.3 & Q3.0 & 99 & . & YES \\
HD332755 & O7.5Ib-II & 54 & 110 & 34.8 ± 1.4 & 3.37 ± 0.21 & 4.20 ± 0.23 & 0.18 ± 0.08 & 25 & -12.5 & Q3.0 & 4 & LS & NO \\
HD225160 & O8Iabf & 77 & 103 & 32.9 ± 0.5 & 3.30 ± 0.10 & 4.17 ± 0.10 & 0.13 ± 0.03 & 21 & -12.3 & Q2.0 & 5 & . & NO \\
BD-114586 & O8Ib(f) & 74 & 68 & 32.7 ± 1.0 & 3.21 ± 0.13 & 4.22 ± 0.15 & 0.15 ± 0.05 & 22 & -12.4 & Q2.0 & 2 & SB1? & NO \\
HD125241 & O8.5Ib(f) & 118 & 95 & 32.2 ± 0.9 & 3.27 ± 0.11 & 4.14 ± 0.13 & 0.17 ± 0.06 & 15 & -12.5 & Q2.2 & 4 & LPV/ & YES \\
HD61347 & O9Iab & 96 & 101 & 30.7 ± 0.7 & 3.12 ± 0.05 & 4.23 ± 0.05 & 0.14 ± 0.04 & 30 & -12.5 & Q3.0 & 2 & . & . \\
HD148546 & O9Iab & 85 & 95 & 31.8 ± 0.7 & 3.22 ± 0.10 & 4.16 ± 0.11 & 0.132 ± 0.025 & 30 & -12.4 & Q3.0 & 4 & LS & YES \\
HD173783 & O9Iab & 89 & 90 & 31.4 ± 0.8 & 3.02 ± 0.04 & 4.33 ± 0.04 & 0.20 ± 0.08 & 13 & -12.7 & Q3.0 & 4 & . & . \\
HD151018 & O9Ib & 67 & 65 & 32.6 ± 0.8 & 3.33 ± 0.05 & 4.09 ± 0.05 & 0.14 ± 0.05 & 13 & -12.4 & Q3.0 & 1 & . & NO \\
HD123008 & ON9.2Iab & 62 & 94 & 31.2 ± 0.9 & 3.05 ± 0.07 & 4.27 ± 0.07 & 0.14 ± 0.04 & 30 & -12.5 & Q3.0 & 2 & . & NO \\
HD226868 & O9.7Iabpva & 95 & 70 & 28.9 ± 0.6 & 3.01 ± 0.07 & 4.21 ± 0.07 & 0.14 ± 0.06 & 10 & -12.5 & Q3.0 & 14 & SB1 & YES \\
HD225146 & O9.7Iab & 67 & 88 & 28.6 ± 0.9 & 3.11 ± 0.12 & 4.08 ± 0.14 & 0.16 ± 0.08 & 14 & -12.6 & Q3.0 & 5 & LS & YES \\
HD75222 & O9.7Iab & 86 & 65 & 30.2 ± 0.6 & 3.17 ± 0.09 & 4.10 ± 0.11 & 0.13 ± 0.03 & 3 & -12.5 & Q3.0 & 5 & LS & YES \\
HD14442 & O5n(f)p & 324 & 0 & 39.6 ± 1.5 & 3.84 ± 0.12 & 3.93 ± 0.14 & 0.14 ± 0.04 & 25 & -12.5 & Q2.3 & 15 & LS & NO \\
HD117797 & O7.5fp & 150 & 80 & 33.7 ± 0.8 & 3.32 ± 0.05 & 4.18 ± 0.05 & 0.151 ± 0.022 & 30 & -12.3 & Q2.2 & 4 & LPV & NO \\
     \end{longtable}
}


\end{document}